\documentclass[a4paper,aps,reprint,pra, 11pt, onecolumn, superscriptaddress, nofootinbib, notitlepage]{revtex4-1}
\usepackage[utf8]{inputenc}
\usepackage{amsmath}
\usepackage{amssymb}
\usepackage{graphicx}
\usepackage{epstopdf}
\usepackage{inputenc}
\usepackage{amsfonts}
\usepackage{calrsfs}
\usepackage{multimedia}
\usepackage{subfig}
\usepackage[cal=boondox]{mathalfa}
\usepackage{caption}
\usepackage{tensor}
\usepackage{amsmath}
\usepackage{physics}
\usepackage{tcolorbox}
\usepackage{tensor}
\usepackage{bbm}
\usepackage{dirtytalk}

\newcommand{\Reals}{\mathbb{R}}

\makeatletter
\def\p@subsection{}
\makeatother

\begin{document}

\begin{sloppypar}
\title{An overview of the gravitational spin Hall effect}

\author{Marius A. Oancea}
\email{marius.oancea@aei.mpg.de}
\affiliation{Max Planck Institute for Gravitational Physics (Albert Einstein Institute), Am M\"uhlenberg 1, D-14476 Potsdam, Germany}

\author{Claudio F. Paganini}
\email{claudio.paganini@monash.edu}
\affiliation{School of Mathematical Sciences, 9 Rainforest Walk, Monash University, Victoria 3800, Australia}

\author{J\'er\'emie Joudioux}
\email{jeremie.joudioux@aei.mpg.de}
\affiliation{Max Planck Institute for Gravitational Physics (Albert Einstein Institute), Am M\"uhlenberg 1, D-14476 Potsdam, Germany}

\author{Lars Andersson}
\email{lars.andersson@aei.mpg.de}
\affiliation{Max Planck Institute for Gravitational Physics (Albert Einstein Institute), Am M\"uhlenberg 1, D-14476 Potsdam, Germany}

\begin{abstract}

In General Relativity, the propagation of electromagnetic waves is usually described by the vacuum Maxwell's equations on a fixed curved background. In the limit of infinitely high frequencies, electromagnetic waves can be localized as point particles, following null geodesics. However, at finite frequencies, electromagnetic waves can no longer be treated as point particles following null geodesics, and the spin angular momentum of light comes into play, via the spin-curvature coupling. We will refer to this effect as the gravitational spin Hall effect of light. Here, we review a series of theoretical results related to the gravitational spin Hall effect of light, and we compare the predictions of different models. The analogy with the spin Hall effect in Optics is also explored, since in this field the effect is well understood, both theoretically and experimentally. 
\end{abstract}

\maketitle

\tableofcontents

\section*{Introduction}

In General Relativity, the motion of free falling particles is described by causal geodesics. A priori, this motion does not take into account the internal structure of these particles. However, it should be expected that the internal structure, such as the spin degree of freedom, has an influence on the motion, as is the case in other fields, such as Optics and Condensed Matter Physics. In General Relativity, attempts to describe how this internal structure corrects the motion of a body has, for instance, been addressed by Mathisson, Papapetrou, and Dixon, in their celebrated equations. Nonetheless, there exist numerous examples where the spin degree of freedom affects the dynamics of fields and particles. One such important effect is the spin Hall effect. 

\subsection*{Spin Hall effects}

In Condensed Matter Physics, the spin Hall effect (SHE) of electrons was first predicted in 1971 \cite{originalSHE1,originalSHE2}, and describes the appearance of a spin current, transverse to the electric charge current propagating in a material. The effect was first observed by Bakun et al. in 1984 \cite{originalSHE3} as the inverse spin Hall effect, and only later on, in 2004, was the direct spin Hall effect observed in semiconductors \cite{originalSHE4}. The source of this effect is the relativistic spin-orbit coupling between a particle's spin and its center of mass motion inside a potential. Detailed reviews about the SHE of electrons can be found in \cite{SHE_review, SHE_review1}.

A similar effect, called the spin Hall effect of light (SHE-L), is present in the case of electromagnetic waves propagating inside an inhomogeneous optical medium. In this case, the spin-orbit coupling comes from the interaction of the polarization degree of freedom with the gradient of the refractive index of the medium, resulting in a transverse shift of the wave packet motion, in a direction perpendicular to the gradient of the refractive index. The first known forms of a SHE-L are the Goos--H\"anchen effect \cite{GH_effect}, originally reported in 1947, and the Imbert-–Fedorov effect \cite{fedorov2013theory,imbert1972calculation}, reported in 1955. These effects involve polarization-dependent transverse shifts of light beams undergoing refraction or total internal reflection. A recent review of these effects can be found in \cite{Bliokh2013}. Later on, polarization-dependent propagation of light inside an inhomogeneous optical medium was reported under the name ``optical Magnus effect" \cite{OpticalMagnus,SHE-L_original}, in analogy with the Magnus effect experienced by spinning objects moving through a fluid.  This was followed by the work of Onoda et al. \cite{SHE_original} (who introduced the term ``Hall effect of light"), Bliokh et al. \cite{Bliokh2004,Bliokh2004_1,Bliokh2008} and Duval et al. \cite{Duval2006,Duval2007,Duval2013}. The first experimental observation of the SHE-L came in 2008 \cite{SHEL_experiment, Bliokh2008}. Reviews about the current state of the research can be found in \cite{SOI_review,SHEL_review}.

\subsection*{Gravitational spin Hall effect}

The purpose of this paper is to review the existing attempts to describe a \emph{gravitational spin Hall effect} (G-SHE). 
Considering the dynamics of a localized wave packets or a spinning particle, by G-SHE we mean any spin dependent correction of this dynamics, in comparison to the dynamics of a scalar field or geodesic motion. This should extend to General Relativity the spin Hall effects known from Condensed Matter Physics and Optics. The role of the inhomogeneous medium is now played by spacetime itself, and the spin-orbit coupling is a consequence of the interaction between the spin degree of freedom and the curvature of spacetime. This effect is expected to be present for all spin-fields (some examples are the Dirac field, electromagnetic waves and linear gravitational waves) propagating in a non-trivial, fixed spacetimes. Throughout this review, our main focus will be on the G-SHE of light, since the corresponding effects for other spin-fields are similar, and most of the relevant literature focuses on the propagation of light, and electromagnetic waves in general. 

One motivation for studying the G-SHE comes from the fact that electromagnetic waves propagating in curved spacetimes are formally described by the same set of equations as electromagnetic waves propagating inside some optical medium, in flat spacetime. The properties of the optical medium can be related to the components of the metric tensor describing the curved spacetime. The experimental observation of the SHE-L in inhomogeneous optical media, together with the correspondence to curved spacetimes, suggests that this effect might as well play a role for waves propagating in curved spacetimes, in which context it is usually neglected. It is conceivable that the G-SHE might have experimentally observable consequences, for example, in the form of corrections to gravitational lensing.

However, in curved spacetime new conceptual difficulties emerge. For example, in inhomogeneous optical media, the modification of the ray trajectories of light does not present a problem for the theory, as light is slowed down, and a modified trajectory is expected to remain well within the domain permitted by causality. On the other hand, when we are looking at the trajectories of light in curved spacetimes, we are usually talking about null geodesics. If we consider modifications of the null geodesics by the G-SHE, we have to keep in mind that light beams are approximate solutions to Maxwell's equations, and that Maxwell's equations do respect the universal speed limit. 

Various approaches have been proposed in the literature to describe the G-SHE. Based on the used methods, they can be grouped as follows:

\begin{itemize}
    \item The Mathisson--Papapetrou--Dixon (MPD) equations, or their equivalent form for massless particles, the Souriau--Saturnini (SouSa) equations. These are based on a multipole expansion around a trajectory. The discussion will be mainly based on the work of Souriau \cite{souriau1974modele}, Saturnini \cite{saturnini1976modele}, and Duval et al. \cite{Duval,Duval2018hzh}, since these are the main approaches for the massless case. 
    
    \item What we will refer to as the quantum mechanical approach, which is based on methods adapted from Relativistic Quantum Mechanics, such as the Foldy--Wouthuysen transformation, and semiclassical Hamiltonians with Berry phase terms. We will mainly focus on the results of Gosselin et al. \cite{SHE_QM1,SHE_QM2}. 
    
    \item The geometrical optics approach, which is based on next to leading order corrections in the WKB expansion\footnote{We will use the terms WKB approximation, eikonal approximation, geometrical optics approximation, Gaussian beam approximation as more or less synonymous. Some authors differentiate between them based on the number of terms retained or whether the phase function is real or complex. However, to us, these distinctions seem to be inconsistent in the literature and as far as we are aware the preference for one or the other names is just depending on the different communities. Therefore we decided to use them interchangeably.}. The main focus will be on the modified geometrical optics approach proposed by Frolov et al. \cite{Frolov}, and later developed by Yoo \cite{covariantSpinoptics}, and Dolan \cite{spinorSpinoptics,spinorSpinoptics2}. 
\end{itemize}

However, little has been said about how these different approaches relate to each other, despite the fact that they do not all arrive at the same results. The goal of this paper is to collect these results, and provide a systematic picture of the existing G-SHE theory. Since some of these theoretical models have contradictory predictions, it is important to understand the assumptions and motivation behind each model, as well as their limitations. Even though we are currently far from a complete understanding of the G-SHE, our hope is that this discussion will serve as a starting point, ultimately leading towards a deeper understanding of the effect.

Finally, we would like to emphasize that the effect under consideration here is purely classical. Quantum electrodynamics corrections to the trajectories of photons, such as considered in \cite{daniels1994faster,shore2003quantum,lafrance1995gravity,chen2015strong,chen2017strong,shore2002faster,shore2001accelerating,shore1996faster}, that violate causality \cite{shore2006causality,konstantinov1998superluminal,dolgov1998superluminal,shore1996faster}, will not be the subject of our discussion.

\subsection*{Overview}
We start in section \ref{sec:inhomo} with a discussion about the SHE-L in inhomogeneous optical media, which has been well studied both theoretically and experimentally. In particular, we discuss the different types of angular momentum of light, the Berry phase, and the equivalence between Maxwell's equations in curved spacetimes, and Maxwell's equations inside some optical medium. We also present a derivation of the SHE-L equations of motion, and discuss the existing experimental results. In section \ref{sec:spinningparticles}, we discuss the MPDT and the SouSa equations, and present some of the known theoretical predictions. In section \ref{sec:quantum}, we introduce the quantum mechanical approach. In section \ref{sec:maxwell}, we discuss the geometrical optics approximation for Maxwell's equations. In section \ref{sec:MPDequivalence} we present the known equivalence between geometrical optics and the linearized MPDT equations, as well as the equivalence between the quantum mechanical approach and the linearized MPDT equations for massive particles. Finally, in section \ref{sec:discussion} we will discuss the relation between the different approaches towards the G-SHE.

\section{SHE-L in Inhomogeneous Optical Media}\label{sec:inhomo}

In this section, we briefly present some basic features of the SHE-L in inhomogeneous optical media. At first glance this may appear disconnected from our main goal of investigating effects in curved spacetime. However, the concepts and methods described in this section are expected to apply in a General Relativistic context as well. We believe that the development and understanding of the G-SHE will benefit from analogies with Optics and Condensed Matter Physics, where the theory is in a more mature state, and SHEs have been experimentally observed.

We start by discussing the different types of the angular momentum that electromagnetic waves can carry, and how the spin-orbit interactions of light result from the conservation of the total angular momentum. Next, the notion of the Berry phase is introduced, and its relation to the SHE-L is explained. A derivation of the SHE-L equations of motion is presented, based on the work of Ruiz and Dodin \cite{Ruiz2015}. We believe this to be a transparent derivation, showing how the SHE-L arises from Maxwell's equations, without having to introduce any Quantum Mechanical notions. We will close this section by discussing the connection between Maxwell's equations in curved spacetime and Maxwell's equations in flat spacetime, in the presence of an inhomogeneous optical medium. This will serve as one of the main motivations for studying SHEs in curved spacetime. A more extended presentation of the SHE-L can be found in \cite{SOI_review, SHE_review} and references therein.

\subsection{Angular Momentum of Light}\label{sec:angular}

It is well known that electromagnetic waves can carry angular momentum \cite{jackson}. Following classical Maxwell's theory, the angular momentum density is given by the cross product of position vector $\mathbf{x}$ with the Poynting vector $\mathbf{E} \cross \mathbf{B}$. The total angular momentum of the electromagnetic field is the space integral of this quantity \cite{jackson}:

\begin{equation}
    \mathbf{J} = \epsilon_0 \int \mathbf{x} \cross \left( \mathbf{E} \cross \mathbf{B}  \right)dx^3 ,
\end{equation}
where $\epsilon_0$ is the vacuum permittivity. Furthermore, the total angular momentum can be split into two parts:

\begin{equation} \label{eq:light_AM}
    \mathbf{J} = \mathbf{S} + \mathbf{L} = \epsilon_0 \int \left( \mathbf{E} \cross \mathbf{A}  \right)dx^3 + \epsilon_0 \sum_{i = 1}^3 \int E_i \left( \mathbf{x} \cross \nabla  \right) A_i dx^3 .
\end{equation}

The first term, $\mathbf{S}$, represents the spin angular momentum, and can be associated with the polarization of the electromagnetic wave. The second term, $\mathbf{L}$, represents the orbital angular momentum, and was mostly ignored until the early 1990s, when it was shown that Laguerre--Gaussian light beams carry well defined spin and orbital angular momentum \cite{Allen92}. Detailed reviews about how the angular momentum of light shaped the last 25 years of developments in the science of light, covering both theoretical and experimental ground, can be found in \cite{AM_Light,AM_Light2, lightAM_review}.

When considering the propagation of light in inhomogeneous optical media, it is convenient to adopt the paraxial beam approximation. This means that the considered electromagnetic wave packet does not spread significantly during its propagation, so it can effectively be described by a ray trajectory. Within this approximation, considering a beam with mean wave vector $\mathbf{P}$ (and $P = \abs{\mathbf{P}}$), the total angular momentum of light can be split into three distinct components \cite{SOI_review, lightAM_review}:

\begin{itemize}
  \item Spin angular momentum (SAM): this corresponds to the first term in equation \eqref{eq:light_AM}, and it is related to the polarization of electromagnetic waves. The SAM per photon can take values $\sigma = \pm \hbar$, and in flat spacetime it is aligned with the direction of propagation of the beam: 
\end{itemize} 
\begin{equation}
    \mathbf{S} = \sigma \frac{\mathbf{P}}{P}.
\end{equation}

\begin{itemize}  
  \item Intrinsic orbital angular momentum (IOAM): this is characteristic for electromagnetic beams with helical wavefronts, such as Laguerre--Gaussian \cite{Allen92}, Bessel \cite{Bessel} or exponential beams \cite{Exponential}. Beams with IOAM are generally described by a topological charge $\mathcal{l}$, which represents the twisting degree of the wavefronts. The IOAM per photon can take any integer value $\mathcal{l} = 0, \pm \hbar, \pm 2\hbar, ...$, and in flat spacetime it is aligned with the direction of propagation of the beam:
\end{itemize}
\begin{equation}
    \mathbf{L_{int}} = \mathcal{l} \frac{\mathbf{P}}{P}.
\end{equation}

\begin{itemize}
  \item Extrinsic orbital angular momentum (EOAM): this is in direct analogy with the mechanical angular momentum for massive particles, and it is present for beams propagating at a distance from the origin of the coordinate system (the origin might correspond to some special point of an applied external potential). The EOAM is given by the cross product of the centroid of the propagating beam, $\mathbf{R}$, and its momentum, $\mathbf{P}$:
\end{itemize}
\begin{equation}
    \mathbf{L_{ext}} = \mathbf{R} \cross \mathbf{P}.
\end{equation}

The second term in equation \eqref{eq:light_AM} is the sum of the IOAM and EOAM. Thus, the total angular momentum of paraxial light beams can be written as:

\begin{equation}
    \mathbf{J} = \mathbf{S} + \mathbf{L} = \mathbf{S} + \mathbf{L_{int}} + \mathbf{L_{ext}}.
\end{equation}

The conservation of the total angular momentum will induce the spin-orbit interactions of light, resulting in the SHE-L and other related effects. For example, if we consider a system where only SAM and EOAM are present, the conservation of the total angular momentum will induce the SHE-L. Another possible example is a system with IOAM and EOAM, where conservation of the total angular momentum will result in a similar effect, called the orbital Hall effect \cite{Bliokh2006, SOI_review}. In particular, IOAM plays a special role since the topological charge $\mathcal{l}$ can take any integer value, thus one can in principle prepare beams that carry significant amounts of angular momentum. Optical beams with IOAM up to $10^4 \hbar$ per photon have been reported \cite{highOAM}. 

Also, the discussion presented here is not limited to electromagnetic waves. The same splitting of the total angular momentum can be considered for any other spin-field, and conservation of the total angular momentum will give raise to the corresponding spin-orbit interactions. In particular, it is worth emphasizing that electrons carrying IOAM are attracting a lot of attention \cite{IOAM_electrons1,IOAM_electrons2,IOAM_electrons3,IOAM_electrons4,IOAM_electrons5}, and gravitational waves carrying IOAM have also been theoretically studied in \cite{GW_IOAM1,GW_IOAM2,GW_IOAM3,GW_IOAM4}. 

\subsection{Berry Phase}\label{sec:berry}

The Berry phase plays a central role in the description of SHEs, both in Optics \cite{SHE_original, SOI_review, Bliokh2009}, and in Condensed Matter Physics \cite{SHE_QM2,Murakami2006,Sinitsyn2007,Berry_electronic}. For example, by considering relativistic wave equations, such as the Dirac equation or Maxwell's equations, the evolution of the spin degree of freedom will be influenced by the Berry phase, while the spin-orbit coupling will imprint the effect of the Berry phase on the corresponding point-particle equations of motions, resulting in a SHE.

As originally described by Michael Berry \cite{Berry_original}, the adiabatic evolution of a quantum system changes the wavefunction by an additional phase factor, referred to as Berry phase or geometrical phase. The quantum system is considered to remain in some $n$th eigenstate of the Hamiltonian $\hat{H}(\mathbf R)$: 

\begin{equation}
    \hat{H}(\mathbf R) \ket{ \Psi_n (\mathbf R) } = E_n (\mathbf R) \ket{ \Psi_n (\mathbf R) },
\end{equation}
where $\mathbf R=\mathbf R(t)$ represents the set of parameters varying adiabatically. The adiabatic evolution of the parameters is considered in the sense of Kato \cite{Kato1950}, and it will define a parallel transport of the wavefunction along the path in parameter space \cite{Berry_book}. A well known example of such a system is a spin-$\frac{1}{2}$ particle in a slowly changing magnetic field $\mathbf{B}(t)$ \cite{Berry_book}. In this case, the set of parameters $\mathbf R(t)$ is identified with the magnetic field $\mathbf{B}(t)$, and for magnetic fields of constant magnitude the parameter space will have $S^2$ topology. 

When the parameters $\mathbf R$ vary along a closed loop $C$ in parameter space, such that $\mathbf R(0) =\mathbf  R(T)$, the wavefunction acquires an additional Berry phase $\gamma_n(C)$:

\begin{align}
    \ket{\Psi_n (\mathbf R(T))} = e^{i \gamma_n(C)} e^{-\frac{i}{\hbar} \int_{0}^{T} E_n(\mathbf R(t))dt } \ket{\Psi_n (\mathbf R(0))}, \\ 
    \gamma_n(C) =i \oint_C \bra{\Psi_n (\mathbf R)} \nabla_\mathbf R \ket{\Psi_n (\mathbf R)} \cdot d\mathbf R = \oint_C \mathbf{A_R} \cdot d\mathbf{R}.
\end{align}

The Berry phase can be expressed in terms of the Berry vector potential, $\mathbf{A_R}$, also called the Berry connection. Furthermore, if we consider an arbitrary hypersurface in parameter space, such that $\partial \Sigma = C$, and by using Stokes' theorem, we can rewrite the Berry phase as:

\begin{equation}
  \gamma_n(C) = \int_\Sigma \nabla \cross \mathbf{A_R} \cdot d\mathbf{S} = \int_\Sigma \mathbf{F_R} \cdot d\mathbf{S}.  
\end{equation}

In the above expression $\mathbf{F_R}$ is called the Berry curvature, since it describes the geometrical properties of the parameter space. In analogy with classical electrodynamics, we can think of $\mathbf{A_R}$ as a ``magnetic" vector potential, and of $\mathbf{F_R}$ as the corresponding ``magnetic" field in the parameter space. Then, one can regard the Berry phase $\gamma_n(C)$ as the flux of $\mathbf{F_R}$ through the surface $\Sigma$ \cite{Berry_book}.

Shortly after Berry's original paper, an elegant mathematical formulation was introduced by Barry Simon, who represented the geometrical phase factor by the holonomy of a connection on a Hermitian line bundle \cite{Berry_Simon}. Later on, generalizations of the Berry phase were introduced by Wilczek and Zee for systems with degenerate spectra \cite{Wilczek-Zee}, and by Aharonov and Anandan for systems undergoing general cyclic evolution, that is not necessarily adiabatic \cite{Aharonov-Anandan,Anandan1988}. Extensions for noncyclic evolution exist as well \cite{Berry_noncyclic1,Berry_noncyclic2,Berry_noncyclic3}. 

From the definition of the Berry phase presented above, one might conclude that this is a purely Quantum Mechanical effect, and it should not be present at the level of classical theories. However, as it can be seen from \cite{classical_Berry1,classical_Berry2}, the Berry phase naturally occurs in classical field theories as well.

Generally, the study of SHEs involves the propagation of localized wave packets inside some inhomogeneous medium. Nevertheless, it is instructive to look at the following basic example. If we consider electromagnetic waves described by classical Maxwell's equations, we can easily see how the Berry phase arises naturally, without considering any Quantum Mechanical effects \cite{Maxwell_Berry1,Maxwell_Berry2,Maxwell_Berry3}. The intrinsic topological structure of Maxwell's equations in vacuum is revealed as soon as one performs a plane wave expansion for the electromagnetic waves. Using this description, electromagnetic waves are characterized by a wave vector $\mathbf{k}$ and a complex polarization vector $\mathbf{e(k)}$, together with the transversality condition $\mathbf{k} \cdot \mathbf{e(k)} = 0$. Furthermore, the space of possible wave vectors is constrained by the dispersion relation (also called on-shell condition) $\abs{\mathbf{k}}^2 = \omega^2(\mathbf{k})$, which implies that the $\mathbf{k}$-space will have $S^2$ topology \cite{Maxwell_Berry2}. The polarization vectors $\mathbf{e(k)}$ form a 2-dimensional complex vector space, and due to the transversality condition they will lie in a tangent plane to the spherical space of $\mathbf{k}$ vectors.

By identifying the parameter space from the standard treatment of the Berry phase with the $\mathbf{k}$-space of electromagnetic waves, one can see how the Berry phase arises at the classical level \cite{Haldane1986,Haldane1987}. Considering an electromagnetic wave that follows a closed loop in $\mathbf{k}$-space, the polarization vector $\mathbf{e(k)}$ will be parallel transported around this loop, and, due to the curvature of the $\mathbf{k}$-space, it will get rotated by a geometrical phase factor proportional to the solid angle enclosed by the loop \cite{Berry_book} (a  visual example of this process is also presented in \cite{Maxwell_Berry3}). This rotation of the polarization vector was already known in 1938, when it was investigated by Rytov \cite{rytov1938}, followed by the work of Vladimirskii \cite{vladimirskii1941} (for this reason, the effect is generally referred to as Rytov or Rytov--Vladimirskii rotation). The effect was experimentally observed for the first time in 1984 by Ross \cite{Ross1984}, followed by the work of Chiao, Tomita and Wu \cite{Chiao-Wu,Tomita-Chiao}.

Even though it will not be considered in the present review, a similar effect, called the Pancharatnam phase, will also arise if the polarization state space is identified as the parameter space and adiabatic evolution of the polarization vector is considered \cite{Pancharatnam1956,Berry-Pancharatnam}. This effect was also observed experimentally \cite{Pancharatnam-experiment}.

However, when it comes to curved spacetime, there are few theoretical studies discussing the Berry phase, and no experimental results. A first study of the Berry phase for waves propagating in a weak gravitational field was presented in \cite{Berry_CS1}, and further developed by several authors \cite{Berry_CS2,Berry_CS3,Berry_CS4,Berry_CS5,Berry_CS6,Berry_CS7,Palmer2012}. In some of the previously mentioned papers the Berry phase goes by the name ``Wigner rotation" or ``Wigner phase", but this is just a difference in terminology, arising mainly from the connection with Wigner's little group for massless particles \cite{Berry_Wigner}. Even though there is no experimental observation of geometric phases in curved spacetime, there is a recent experimental proposal for measuring the Wigner phase of photons in the gravitational field of the Earth, with a predicted phase difference that could in principle be measured with currently available technology \cite{kohlrus2018}.

\subsection{SHE-L Equations of Motion}\label{sec:SHEL_eq}

The SHE-L in inhomogeneous optical media can be viewed as a consequence of the spin-orbit coupling between SAM and EOAM, resulting in the helicity dependence of the ray trajectories. In terms of the Berry phase, the SHE-L can be described by considering $\mathbf{k}$-space as parameter space. Then the Berry curvature of $\mathbf{k}$-space will act as a ``Lorentz force" on ``charged" particles, where the ``charge" will be represented by the helicity of photons. Thus, the SHE-L can be viewed as a consequence of Berry curvature in momentum space \cite{SHE_original}. 

The point-particle equations of motion describing the SHE-L have been obtained by different authors, using different methods. These include postulating an effective ray Lagrangian or Hamiltonian \cite{SHE_original}, using geometrical optics with a modified eikonal ansatz on Maxwell's equations \cite{Bliokh2004,Bliokh2004_1}, or considering a mechanical model for photons, as inspired by the description of spinning particles in General Relativity \cite{Duval2006}. 

However, due to the variety of these different methods, the connection between the SHE-L and Maxwell's equations is not always clear. In order to remove any possible source of confusion, in this section we will review the derivation of the SHE-L in inhomogeneous optical media, as presented by Ruiz and Dodin \cite{Ruiz2015}. Their approach is based on a first-principle variational formulation of the geometrical optics approximation for Maxwell's equations, and reproduces the previously known results of Zel'dovich \cite{SHE-L_original,OpticalMagnus}, Onoda \cite{SHE_original}, Bliokh \cite{Bliokh2004,Bliokh2004_1} and Duval \cite{Duval2006}, without postulating ray Lagrangians or introducing ad hoc modifications of the eikonal ansatz. Then, the SHE-L readily follows from Maxwell's equations, and the classical nature of the effect becomes apparent. Notably this method can also be applied to other field equations \cite{Dodin2014,Ruiz2015(2),Ruiz2015(3),Ruiz2017,Ruiz2017(2),Dodin2018}.

Ruiz and Dodin start by considering electromagnetic waves propagating in an isotropic dielectric medium (the case of more general dispersive media can be found in \cite{Ruiz2017}). In this case, the electric and magnetic fields are described by the following equations:

\begin{align}
     \partial_t \mathbf{E} & = \frac{c}{\varepsilon} \nabla \cross \mathbf{H}, \\
      \partial_t \mathbf{H} & = -\frac{c}{\mu} \nabla \cross \mathbf{E},
\end{align}
where $c$ is the speed of light in vacuum, $\varepsilon = \varepsilon(\mathbf{x})$ is the electric permittivity, and $\mu=\mu(\mathbf{x})$ is the magnetic permeability. Following \cite{Ruiz2015}, we can introduce the normalized fields $\mathbf{\Bar{E}} = \sqrt{\varepsilon} \mathbf{E}$ and $\mathbf{\Bar{H}} = \sqrt{\mu} \mathbf{H}$, in order to cast the field equations in the form:

\begin{align}
    \partial_t \mathbf{\bar{E}} & = \frac{c}{n} \nabla \cross \mathbf{\Bar{H}} - \frac{c}{n}\nabla (\ln \sqrt{\mu}) \cross \mathbf{\bar{H}}, \\
    \partial_t \mathbf{\bar{H}} & = -\frac{c}{n} \nabla \cross \mathbf{\Bar{E}} - \frac{c}{n}\nabla (\ln \sqrt{\varepsilon}) \cross \mathbf{\bar{E}},
\end{align}
where $n = \sqrt{\varepsilon \mu}$ is the refractive index of the medium. Note that the second terms in the above equations are proportional to the first order derivatives of the medium properties and thereby are small, albeit not negligible.

It is well known that Maxwell's equations can be cast in a Schr\"{o}dinger form, and various formulations have been proposed by different authors \cite{Birula_wavefunction1, Birula_wavefunction2, Maxwell_Berry3, NC_Maxwell,EM_Schrodinger}. In the present case, following \cite{Ruiz2015}, the above equations can be rewritten in the following way:

\begin{equation} \label{eq:Maxwell_Schrodinger}
    i \partial_t \psi = H \psi,
\end{equation}
where the vector wavefunction $\Psi$ has $6$ components: 

\begin{equation}
    \psi (\mathbf{x},t) = 
    \begin{pmatrix} 
    \mathbf{\bar{E}}   \\
    \mathbf{\bar{H}}   \\
   \end{pmatrix} ,
\end{equation}
and the Hamiltonian is a $6 \cross 6$ matrix:

\begin{equation}
    H(\mathbf{x}, \mathbf{\hat{k}}) = \frac{c}{n} \pmb{\lambda} \cdot \mathbf{\hat{k}} + \mathcal{A},
\end{equation}
where $\mathbf{\hat{k}} = -i \nabla$ is the momentum operator, $\pmb{\lambda}$ are $6 \cross 6$ Hermitian matrices:

\begin{align}
    \pmb{\lambda} & = \begin{pmatrix} 
        0 & i \pmb{\alpha}   \\
        -i \pmb{\alpha}  & 0  \\
         \end{pmatrix}, \\
   \alpha_x = \begin{pmatrix} 
    0 & 0 & 0  \\
    0 & 0 & -i  \\
    0 & i & 0  \\
   \end{pmatrix}, \qquad
    \alpha_y & = \begin{pmatrix} 
    0 & 0 & i  \\
    0 & 0 & 0  \\
    -i & 0 & 0  \\
   \end{pmatrix}, \qquad
   \alpha_z = \begin{pmatrix} 
    0 & -i & 0  \\
    i & 0 & 0  \\
    0 & 0 & 0  \\
   \end{pmatrix},
\end{align}
and the matrix $\mathcal{A}$ has the following form:

\begin{equation}
    \mathcal{A} = \begin{pmatrix} 
    0 & -\pmb{\alpha} \cdot \nabla(\ln \sqrt{\mu})   \\
    \pmb{\alpha} \cdot \nabla(\ln \sqrt{\varepsilon})  & 0  \\
   \end{pmatrix}.
\end{equation}

As described in \cite{Ruiz2015}, equation \eqref{eq:Maxwell_Schrodinger} can be obtained from a variational formulation, with the action:

\begin{equation} \label{eq:action}
    S = \int \mathcal{L} d^4 x,
\end{equation}
where the Lagrangian density, $\mathcal{L}$, takes the following Dirac-like form:

\begin{equation} \label{eq:Lagrangian}
    \mathcal{L} = \frac{i}{2} [\psi^\dagger \gamma^\mu (\partial_\mu \psi) - (\partial_\mu \psi^\dagger) \gamma^\mu \psi] + \psi^\dagger \mathcal{M} \psi.
\end{equation}
The gamma matrices are defined as $\gamma^\mu = (\mathbb{I}_6, c \pmb{\lambda} / n)$, and $\mathcal{M}$ has the following form:

\begin{equation}
    \mathcal{M} = \frac{1}{2 n} \begin{pmatrix} 
    0 & \pmb{\lambda}  \\
    \pmb{\lambda} & 0  \\
   \end{pmatrix} \cdot \nabla \ln \sqrt{\frac{\mu}{\varepsilon}}.
\end{equation}

This variational formulation of Maxwell's equations represents the starting point for the geometrical optics approximation, as described by Ruiz and Dodin in \cite{Ruiz2015}. However, it should be stressed out that having the Lagrangian density written in this particular form is just a matter of convenience and not a strict requirement. An extension of the formalism, that does not rely on any particular form of the Lagrangian density, has been presented in \cite{Ruiz2017}. 

The following eikonal ansatz is considered:

\begin{equation} \label{eq:eikonal}
\psi (t,\mathbf{x})=\xi e^{i \theta / \epsilon}, 
\end{equation}
where $\xi$ is a slowly varying complex amplitude, $\theta$ is a rapid real phase, and $\epsilon$ is a dimensionless expansion parameter. As usual, the length scale $L$ over which the properties of the medium vary significantly is assumed to be large in comparison to the wavelength of the electromagnetic wave:

\begin{equation} \label{eq:lengthscales}
    \epsilon = \frac{1}{\abs{\mathbf{k}} L} \ll 1.
\end{equation}

The wave vector is defined as $\mathbf{k} = \nabla \theta$, and the frequency is $\omega = -\partial_t \theta$. The eigenfrequencies and the corresponding eigenmodes are found from the geometrical optics limit of equation \eqref{eq:Maxwell_Schrodinger}, where only terms of order $\epsilon^0$ are retained. This means that we will neglect the $\mathcal{A}$ term from the Hamiltonian, since this includes first-order derivatives of the medium properties, and  therefore is of order $\epsilon^1$. We are left with the following eigenvalue problem:

\begin{equation}
    H_0 (\mathbf{x}, \mathbf{k}) \xi = \omega \xi,
\end{equation}
where $H_0 (\mathbf{x}, \mathbf{k}) = c \pmb{\lambda} \cdot \mathbf{k} / n $. Since $H_0$ is a $6 \cross 6$  Hermitian matrix, generally there exist six independent eigenvectors $h_q$, which form a complete basis. Two of the eigenvectors will correspond to longitudinal modes, and the other four eigenvectors correspond to transverse modes. Here, we will be interested only in the propagation of transverse electromagnetic modes with positive frequencies $\omega = k/n$, thus we will only consider the following two orthonormal eigenvectors \cite{Ruiz2015}:

\begin{equation}
    h_1 (\mathbf{k}) = \frac{1}{\sqrt{2}} 
    \begin{pmatrix} 
    \mathbf{e_1}   \\
    \mathbf{e_2}   \\
   \end{pmatrix},
   \qquad \qquad
   h_2 (\mathbf{k}) = \frac{1}{\sqrt{2}} 
    \begin{pmatrix} 
    \mathbf{e_2}   \\
    \mathbf{-e_1}   \\
   \end{pmatrix}.
\end{equation}
Note that the vectors $h_{1,2}$ have 6 components and determine a linear polarization basis, while the vectors $\mathbf{e_{1, 2}}$ have 3 components and determine a plane normal to $\mathbf{e_k}=\mathbf{k}/k$. 

By using the six eigenvectors $h_q$, we can expand the complex amplitude as $\xi = h_q \phi^q$, where $\phi^q$ are scalar functions. However, since we are only considering transverse modes with positive frequency, the only active modes will be those corresponding to $h_{1, 2}$, while the other modes can only become exited through the inhomogeneity of the medium. In this case, the active modes $h_{1, 2}$ are of order $\epsilon^0$, while the other modes will be of order $\epsilon^1$ and can be ignored for the purpose of this calculation \cite{Ruiz2015,Ruiz2015(2)}. The complex amplitude $\xi$ can now be written in the following way \cite{Ruiz2015}:

\begin{equation} \label{eq:polarizationansatz4}
    \xi = h_1 \phi^1 + h_2 \phi^2 + O(\epsilon)= \Xi \phi + O(\epsilon),
\end{equation}
where $\phi$ is a $2 \cross 1$ matrix with complex scalar elements:
\begin{equation}
    \phi = \phi (t, \mathbf{x}) = \begin{pmatrix} 
    \phi^1   \\
    \phi^2   \\
   \end{pmatrix},
\end{equation}
and $\Xi$ is a $6 \cross 2$ matrix having $h_{1, 2}$ as columns:
\begin{equation}
     \Xi = \frac{1}{\sqrt{2}}\begin{pmatrix} 
    \mathbf{e_1} & \mathbf{e_2}   \\
    \mathbf{e_2} & -\mathbf{e_1}   \\
     \end{pmatrix}.
\end{equation}

At this point, we are using the basis formed by the polarization vectors of linearly polarized modes, but we can easily move to a circular polarization basis by using the following transformation:

\begin{equation}
    \phi (t, \mathbf{x}) = Q \eta(t, \mathbf{x}) = \frac{1}{\sqrt{2}}
            \begin{pmatrix}
             1 & 1  \\
             i & -i 
            \end{pmatrix} \eta(t, \mathbf{x}).
\end{equation}

As we will see in what follows, the linear polarization basis is useful for investigating the polarization dynamics, while the circular polarization basis has the advantage that the dynamics of right-handed and left-handed circular polarized eigenmodes is decoupled.

The next step is to insert the eikonal ansatz, as defined in equations \eqref{eq:eikonal} and \eqref{eq:polarizationansatz4}, into the Lagrangian density. After introducing a particular frame choice for $\mathbf{e_{1, 2}}(\mathbf{k})$ and moving to a circular polarization basis, the Lagrangian density under the geometrical optics approximation becomes \cite{Ruiz2015}:

\begin{equation} \label{eq:LagrangianGO}
    \mathcal{L} = -\eta^\dagger (\partial_t \theta + k/n)\eta + \frac{i}{2}[\eta^\dagger (d_t \eta)- (d_t \eta^\dagger)\eta] -\eta^\dagger \sigma_z \Sigma(\mathbf{x},\mathbf{k})\eta ,
\end{equation}
where $\sigma_z$ is the usual Pauli matrix,

\begin{align} 
d_t & = \partial_t + \frac{\mathbf{e_k}}{n} \cdot \nabla, \\
\Sigma(\mathbf{x},\mathbf{k}) & = \dot{\mathbf{k}} \cdot \mathbf{A}(\mathbf{k}) = \frac{k}{n^2} \mathbf{A}(\mathbf{k}) \cdot \nabla n, \\
\mathbf{A}(\mathbf{k}) & = \frac{k_z}{k (k_x^2 + k_y^2)} 
            \begin{pmatrix}
              k_y  \\
              -k_x \\
              0
            \end{pmatrix}.
\end{align}

The first term in equation \eqref{eq:LagrangianGO} is of order $\epsilon^0$ and represents the lowest order geometrical optics approximation, while the following two terms are of order $\epsilon^1$ and represent polarization-dependent corrections. By introducing the reparametrization $\eta = z \sqrt{I}$, where $I = \psi^\dagger \psi$ is the intensity of the wave, and $z$ is a unit complex polarization vector ($z^\dagger z = 1$), the Lagrangian can be expressed as:

\begin{equation} \label{eq:LagrangianGO1}
    \mathcal{L} = -I \left [ \partial_t \theta + \frac{k}{n} -\frac{i}{2} \left(z^\dagger (d_t z) - (d_t z^\dagger) z \right) +\Sigma (\mathbf{x},\mathbf{k}) z^\dagger \sigma_z z \right].
\end{equation}

At this point, the above equation still represents a field Lagrangian, with the dynamical variables given by $I$, $\theta$, $z$ and $z^\dagger$. However, since the intensity $I$ is an overall factor, there is a clear way of localizing waves into the point-particle limit. This can be achieved by requiring the intensity $I$ to be sharply localized, and in the point-particle limit approximated with a Dirac delta function: 

\begin{equation} \label{eq:Diracintensity}
    I(t, \mathbf{x}) = \delta^3 (\mathbf{x} - \mathbf{X}(t)),
\end{equation}
where $\mathbf{X}(t)$ will be the location of the point-particle at time $t$.

Inserting equations \eqref{eq:LagrangianGO1} and \eqref{eq:Diracintensity} into the expression of the action, and performing the integration over the spatial coordinates $\mathbf{x}$, one obtains \cite{Ruiz2015}:

\begin{equation}
    S = \int dt d^3 x \mathcal{L} = \int dt L,
\end{equation}
where $L$ is the corresponding point-particle Lagrangian:

\begin{equation}
    L = \int d^3 x \mathcal{L} = \mathbf{P} \cdot \mathbf{\dot{X}} - \frac{c P}{n} + \frac{i}{2} (Z^\dagger \dot{Z} - \dot{Z}^\dagger Z) - \Sigma(\mathbf{X}, \mathbf{P})Z^\dagger \sigma_z Z.
\end{equation}

This is a point-particle Lagrangian, describing the ray dynamics, where the independent variables are $\mathbf{X}(t)$, $\mathbf{P}(t) = \nabla \theta (t, \mathbf{X}(t))$, $Z(t) = z(t, \mathbf{X}(t))$ and $Z^\dagger (t) = z^\dagger(t, \mathbf{X}(t))$. The ray equations are given by the Euler--Lagrange equations:

\begin{align} 
\mathbf{\dot{P}}(t) & =  \frac{c \mathbf{P}}{n P} + (\partial_\mathbf{P} \Sigma) Z^\dagger \sigma_z Z, \label{eq:Pdot} \\ 
\mathbf{\dot{X}}(t) & =  \frac{c P}{n^2} \nabla n - (\partial_\mathbf{X} \Sigma) Z^\dagger \sigma_z Z, \label{eq:Xdot} \\ 
\dot{Z}(t) & =  -i \Sigma \sigma_z Z, \\ 
\dot{Z}^\dagger(t) & = i \Sigma Z^\dagger \sigma_z.
\end{align}

Here, the first terms in equations \eqref{eq:Pdot} and \eqref{eq:Xdot} are of order $\epsilon^0$ and represent the lowest order geometrical optics approximation. The other terms are of order $\epsilon^1$, and determined polarization-dependent corrections to the ordinary ray trajectories. These correction terms represent the SHE-L. If one restricts to rays with either right-hand or left-hand circular polarization, then $\sigma_z Z = \pm Z$, and the ray Lagrangian becomes \cite{Ruiz2015}:

\begin{equation}
    L = \mathbf{P} \cdot \dot{\mathbf{X}} - \frac{c P}{n} \mp \Sigma (\mathbf{X}, \mathbf{P}) \approx \mathbf{P} \cdot \dot{\mathbf{X}} - \frac{c P}{n} \mp \dot{\mathbf{P}} \cdot \mathbf{A}( \mathbf{P}).
\end{equation}

In this case, the Euler--Lagrange equations will give the following equations for the ray trajectories:

\begin{equation} \label{eq:SHE_Ruiz}
    \dot{\mathbf{P}} = \frac{c P}{n^2} \nabla n, \qquad \qquad
    \dot{\mathbf{X}} = \frac{c \mathbf{P}}{n P} \pm \frac{\dot{\mathbf{P}} \cross \mathbf{P}}{P^3}.
\end{equation}

These equations reproduce the previous results on the SHE-L in inhomogeneous media \cite{SHE-L_original,SHE_original,Bliokh2004,Bliokh2006,Bliokh2008,Duval2006,SOI_review} (in some cases the equations appear in a slightly different form, but this is just due to a rescaling of the momentum and time in equation \eqref{eq:SHE_Ruiz} \cite{Ruiz2015}), and were derived without introducing any extra phase factors into the eikonal ansatz. The Berry phase is already encoded in the polarization dynamics, and can be explicitly calculated as \cite{Ruiz2015},

\begin{equation}
    \Theta(t) = \int_0^t \Sigma (\mathbf{X}(t), \mathbf{P}(t)) dt,
\end{equation}
where $\Sigma = \dot{\mathbf{P}} \cdot \mathbf{A}( \mathbf{P})$ represents the Berry connection.

In equation \eqref{eq:SHE_Ruiz}, the second term on the right hand side of $\dot{\mathbf{X}}$ represents the correction term that determines the SHE-L and can be interpreted as a Lorentz force produced by the Berry curvature in momentum space, with the photon helicity acting as a charge \cite{SOI_review}. In the limit of very short wavelengths, $\lambda \rightarrow 0$, the SHE-L is suppressed, and we recover the classical equations of motion for photons in a medium with arbitrary refractive index $n$. The SHE-L becomes more visible as one increases the wavelength, but one should keep in mind that these equations were derived under the assumption that the wavelength is much smaller than the length scale over which the medium properties varies significantly, as presented in equation \eqref{eq:lengthscales}.

The theoretical predictions of equation \eqref{eq:SHE_Ruiz} were first verified in 2008 by Hosten and Kwait \cite{SHEL_experiment}. Their experiment used the technique of quantum weak measurements in order to amplify the small transverse shift coming from the SHE-L. This was followed a few months later by the experiment of Bliokh, Niv, Kleiner and Hasman \cite{Bliokh2008}. In this case, the authors managed to amplify the SHE-L by multiple reflections inside a glass cylinder. Afterwards, the effect was detected by several other groups, using different experimental methods \cite{SHEL_experiment1,SHEL_experiment2,SHEL_experiment3,SHEL_experiment4}. A more detailed account of the experimental results can be found in \cite{SOI_review,SHE_review}.

\subsection{Treating Curved Spacetime as an Effective Inhomogeneous Medium}\label{sec:equivalence}

One of the main motivations for investigating the possibility of a G-SHE of light comes from the fact that electromagnetic waves propagating in a curved spacetime are formally described by the same set of equations as electromagnetic waves in flat spacetime, propagating inside an inhomogeneous optical medium \cite{Plebansky-Maxwell, Birula_wavefunction1, Birula_wavefunction2}. This type of analogy was first recognized by Eddington, who suggested that the gravitational light bending around the Sun could also be obtained if we consider an appropriate distribution of a refractive material \cite{Eddington}. This was later developed by Gordon \cite{Gordon}, and Plebanski \cite{Plebansky-Maxwell}. For a more recent discussion see \cite{Birula_wavefunction1,Birula_wavefunction2}.

Following Plebanski \cite{Plebansky-Maxwell}, a spacetime described by the metric tensor $g_{\mu \nu}$ can be viewed as an effective medium with perfect impedance matching, described by a tensorial permittivity $\epsilon_{i j}$, a tensorial permeability $\mu_{i j}$, and a magnetoelectric coupling vector $\alpha_i$ (here, Latin indices run from 1 to 3):  

\begin{equation}
\epsilon_{i j} = \mu_{i j} = -\sqrt{-\det g} \frac{g_{i j}}{g_{0 0}}, 
 \qquad  \qquad
\alpha_i =-\frac{g_{0 i}}{g_{0 0}}.
\end{equation}

This correspondence is an example of what is called analogue gravity \cite{Analogue_gravity}, where certain properties of a curved spacetime are reproduced in laboratories using other physical systems. Based on this correspondence, and since the SHE-L was predicted and experimentally observed in inhomogeneous optical media, we expect the effect to be present in curved spacetime as well. Several examples supporting this statement will be discussed in the following sections. 

However, this analogy has its limitations and it should be used with care. The main limitation is that it breaks covariance, and simply writing the metric using different coordinates can result in analog materials with completely different properties \cite{cartographic_analog}.

\section{Spinning Particles in the Pole-Dipole Approximation} \label{sec:spinningparticles}

In this section we will discuss the spin-curvature interaction in the pole-dipole approximation for extended test bodies. Since the focus of our review is on the G-SHE of light, we will only touch on the vast literature for massive spinning particles where the results seem of interest to us for the case of massless particles. For an overview of the massive case see \cite{dixon2015new,steinhoff2015spin}. A discussion of the conceptual issues involved when deriving a worldline for an extended spinning body can be found in \cite{van2016world} for the massive case, and \cite{bailyn1977pole,bailyn1981pole} for the massless case.

\subsection{Mathisson--Papapetrou--Dixon--Tulczyjew
Equations}\label{sec:mpd}
The equation for the worldline of a spinning test body in the context of the pole-dipole approximation was first derived by Mathisson \cite{mathisson2010republication} and Papapetrou \cite{papapetrou1951spinning} by integrating the conservation law of the energy momentum tensor $\nabla_\nu T^{\mu\nu}$ for a multipole expansion of the energy momentum tensor $T^{\mu\nu}$. A covariant derivation was first given by Tulczyjew \cite{tulczyjew1959motion} and Dixon \cite{dixon1964covariant}. The latter containing multipole expansions to any order, for that see also \cite{singh2008analytic}. There are many alternative derivations in the literature \cite{singh2008analytic, ramirez2015lagrangian,vines2016canonical,souriau1974modele,barausse2009hamiltonian}. The Hamiltonian formulation for the MPDT equations in \cite{barausse2009hamiltonian}, and the systematic presentation of the Hamiltonian for different orders of spin in  \cite{vines2016canonical} might be interesting, since the SHE-L equations of motion can also be derived from a Hamiltonian formulation. A particularly transparent derivation can be found in section 2 and 3 of \cite{steinhoff2015spin}, and a slightly more general derivation in section V of \cite{dixon2015new}. A more mathematical derivation including a full discussion of the symplectic structure of the phase space of the dynamic variables can be found in \cite{souriau1974modele}, albeit only available in French. For the definition of multipole moments see \cite{dixon1973definition}. 

The MPDT equations have been subject to extensive research, and we will use them as a reference for other derivations of spin-curvature effects. Recent interest is heavily focused on extreme mass-ratio scenarios as sources for gravitational waves, see for example  \cite{khriplovich1996gravitational,porto2011spin,han2010gravitational} and sources therein.

The MPD equations are given by: 

\begin{align}\label{eq:mpd}
    \dot p^\mu & = -\frac{1}{2}R^\mu_{\phantom{\mu}\nu\kappa\lambda}u^\nu S^{\kappa\lambda},\\
    \dot S^{\alpha\beta}& =p^{[\alpha}u^{\beta]}, \label{eq:mpd1}
\end{align}
where $u^\mu$ denotes the four-velocity of the particle, i.e. the timelike unit tangent vector of the worldline $u^\mu u_\mu=-1$, while $p^\mu $ is the total momentum of the particle. Furthermore, $S^{\mu\nu}$ is the totally antisymmetric spin tensor. The system \eqref{eq:mpd}-\eqref{eq:mpd1} has 10 equations for 13 unknowns (3 for $u^\mu$, 4 for $p^\mu$ and 6 for $S^{\mu\nu}$) and is therefore underdetermined. In particular, we are missing an equation that determines $u^\mu$. This is usually fixed with so called spin supplementary conditions (SSC). The most commonly used SSCs are the following ones:

\begin{itemize}
    \item Tulczyjew--Dixon SSC, $S^{\mu\nu}p_\nu=0$
    \item Pirani SSC, $S^{\mu\nu}u_\nu=0$ \cite{pirani2009republication}
    \item Corinaldesi--Papapetrou SSC, $(\partial_t)_\mu S^{\mu\alpha}=0$, for stationary spacetimes. 
\end{itemize}

Note that the worldlines obtained from different SSC do not coincide. They are usually interpreted as different gauge choices for the ``center of mass" of the extended bodies. According to Dixon \cite{dixon2015new}, the Tulczyjew--Dixon SSC, $S^{\mu\nu}p_\nu=0$, is the only SSC that fixes a unique world line for an extended body.  For a review on the effect of the different SSCs and their physical interpretation, see \cite{costa2015center,costa2018spinning}. For the Tulczyjew--Dixon SSC, $m=p^\mu u_\mu$ can be interpreted as the mass, which is constant along the worldline. For the Pirani SSC, the mass is given by $p^\mu p_\mu=\tilde m $, which is, again, conserved along the worldline. For both SSCs, the magnitude of the spin, $s^2= \frac{1}{2}S_{\mu\nu}S^{\mu\nu}$, is constant along the worldlines. It was shown in  \cite{costa2012mathisson} that various choices are in fact physically equivalent. Therefore, choosing a SSC comes down to practicality and personal preferences. 
From equation \eqref{eq:mpd1} and the Tulczyjew--Dixon SSC, the following relation between the total momentum and the four-velocity can be derived: 

\begin{equation}\label{eq:mpdmomentum}
    p^\mu= m u^\mu + \dot S ^{\mu\nu} u_\nu,
\end{equation}
which provides us with an equation to determine $u^\mu$.
For the Tulczyjew--Dixon SSC, we can define the spin vector $S^\mu$ in the following way:

\begin{equation}\label{eq:spinvector}
    S^\mu= \frac{1}{2m}\epsilon^{\mu\nu\kappa\lambda}p_\nu S_{\kappa\lambda}, \qquad S^{\mu\nu}= \frac{1}{m }\epsilon^{\mu\nu\kappa\lambda}p_\kappa S_{\lambda},
\end{equation}
which also satisfies $S^\mu S_\mu=-s^2$. Note that $s^2$ is a constant of motion along the worldline described by the equations \eqref{eq:mpd}, independent of the SSC. Here, $\epsilon^{\alpha\beta\rho\sigma}$ is the totally antisymmetric Levi--Civita tensor, with $\varepsilon^{0 1 2 3} = 1/\sqrt{- \det g}$.

To linearize the MPDT equations in spin, $s$ is treated as a small parameter in the equations, and all terms quadratic in $s$ are omitted. One generally considers $S^\mu\propto s$, or alternatively $S^{\mu\nu}\propto s$. Linearized in spin, the MPDT equations then have the following form:

\begin{align} \label{eq:linmpd}
    \dot p^\mu & \approx -\frac{1}{2}R^\mu_{\phantom{\mu}\nu\kappa\lambda}u^\nu S^{\kappa\lambda}, \\
    \dot S^{\alpha\beta} & \approx 0,
\end{align}
and equivalently 

\begin{equation}\label{eq:spinvectorequation}
    \dot S^{\alpha} \approx 0.
\end{equation}

This form is sometimes referred to as Mathisson--Papapetrou--Pirani (MPP) equations. In this case, we have: 

\begin{equation}
    p^\mu\approx m u^\mu .
\end{equation}

Therefore, the Tulczyjew--Dixon SSC, $S^{\mu\nu}p_\nu=0$, and the Pirani SSC, $S^{\mu\nu}u_\nu=0$, can be satisfied simultaneously. Hence, when the equations are linearized in spin there is less ambiguity with respect to choosing the correct SSC. 
We will return to the MPP equations in later sections (\ref{sec:diracmpd}, \ref{sec:diracmpdwkb}), where we demonstrate that the MPP equations can be derived from field equations, either by a WKB expansion \cite{rudiger,audretsch}, or by a Foldy--Wouthuysen transformation truncated at linear order in  $\hbar$, in order to derive ``quantum" corrections to geodesic motion.  

For readers interested in the application of the MPDT equations in the context of the astrophysically relevant spacetimes, such as the rotating Kerr black holes \eqref{eq:kerrtaubnut}, \cite{hackmann2014motion,semerak1999spinning,kyrian2007spinning,singh2008perturbation,mashhoon2006dynamics} and sources therein might be a good start. We will omit a deeper discussion at this point, as the case of massive trajectories is not our main focus here. The interested reader could also consult \cite{costa2016spacetime} for an extensive collection of sources on the topic. 

Other equations for worldlines of interest in the context of massive spinning bodies have been derived in \cite{van2016world,d2015covariant,van2016spinning,d2016spinning} from a Hamiltonian formulation. The authors start by postulating a phase space consisting of the position coordinate $x^\mu$, the covariant momentum $\pi_\mu$ and the antisymmetric spin tensor $\Sigma^{\mu\nu}$. Then, they postulate antisymmetric Poisson brackets that define a symplectic structure over the phase space. Finally, they choose a Hamiltonian that generates the evolution of the system. The worldline obtained this way is characterized by the fact that the spin tensor $S^{\mu\nu}$ is covariantly constant along the path. At present, it is unclear to us how this approach relates to the Hamiltonian formalism used in \cite{ramirez2015lagrangian,vines2016canonical}. If the equations obtained in this approach are linearized in spin, they correspond to the MPD equations linearized in spin \eqref{eq:linmpd}. The case of massless particles has not yet been worked out in this model.

\subsection{Souriau--Saturnini Equations}\label{sec:ss} 

In this section we discuss the pole-dipole approximation for massless particles. We note, as a preliminary comment, that it has been argued in the literature \cite{saturnini1976modele,Duval,Duval2018hzh, marsot} that there is a problem with the equations \eqref{eq:souriausaturnini} in flat space, where the equations appear to be singular\footnote{It is not entirely clear to us whether the equations \eqref{eq:souriausaturnini} are indeed singular in the limit of flat spacetime. If we make the replacement $R_{\alpha\beta\lambda\nu}\rightarrow \epsilon R_{\alpha\beta\lambda\nu}$, no terms with negative powers of $\epsilon$ appear. Similarly, if we look at Schwarzschild spacetimes \eqref{eq:schwarzschild}, all non-zero components of the Riemann tensor are proportional to the mass $M$}. The equations \eqref{eq:souriausaturnini} have no direct connection to Maxwell's equations. Nevertheless, interesting results have been obtained using this model, hence we include a discussion at this point. 

The fact that the MPDT equations can be adapted for massless particles was first mentioned by Souriau \cite{souriau1974modele}, and then worked out in detail by Saturnini \cite{saturnini1976modele} (both references only available in French). They start with the MPDT equations \eqref{eq:mpd}, and assume the Tulczyjew--Dixon SSC, $S^{\mu\nu}p_\nu=0$, $S^{\mu\nu}\neq 0$, and for the momentum to be null $p^\mu p_\mu=0$. Then, they obtain the following set of equations, to which we will refer to as the Souriau--Saturnini (SouSa) equations:

\begin{align}\label{eq:souriausaturnini}
 u^\mu&= p^\mu +\frac{2}{R_{\alpha\beta\lambda\nu}S^{\alpha\beta}S^{\lambda\nu}} S^{\alpha\mu}R_{\alpha\beta\lambda\nu}S^{\lambda\nu}p^\beta ,\\
    \dot p^\mu&= s \frac{\sqrt{-g} \epsilon^{\alpha\beta\rho\sigma}R_{\alpha\beta\lambda\nu}S^{\lambda\nu}R_{\rho\sigma\gamma\delta}S^{\gamma \delta}}{8 R_{\alpha\beta\lambda\nu}S^{\alpha\beta}S^{\lambda\nu}} p^\mu ,\\
    \dot S^{\mu\nu}&=p^{[\mu}u^{\nu]}, 
\end{align}
where $g$ is the metric determinant.

One point to note is that, according to \cite{saturnini1976modele}, the condition $p^\mu p_\mu =0$, together with the equations \eqref{eq:souriausaturnini}, implies that $u^\mu u_\mu >0$, and hence the ``massless particles" in this model move on spacelike trajectories. We will discuss in section \ref{sec:discussion} in what limited way such spacelike trajectories can be considered as valid results for particle trajectories that emerge from a causally well behaved field equation.

In the following, we discuss some applications of the SouSa equations and claims attained thereof. In his thesis \cite{saturnini1976modele}, Saturnini showed that for a certain choice of initial condition for the spin, a radially ingoing null geodesic would satisfy equation \eqref{eq:souriausaturnini} and hence he argued, as a first physical result of the model \eqref{eq:souriausaturnini}, that the observation of redshift would not change for massless particles with spin. He also observed, in numerical simulations, that for certain initial conditions in Schwarzschild spacetimes, the equations \eqref{eq:souriausaturnini} with different helicities produce trajectories that are symmetric with respect to the plane of a reference null geodesic with zero spin. However, he deemed the effect to be too small to be observable.

In \cite{Duval2007,Duval2013} Duval et al. reproduce the results of Fedorov \cite{fedorov2013theory} and Imbert \cite{imbert1972calculation} for the polarization dependent reflection of light, using the framework introduced by Souriau and Saturnini.

In \cite{Duval}, Duval and Schücker studied the SouSa equations in a Robertson Walker spacetime. By numerically integrating \eqref{eq:souriausaturnini} with a non-zero orthogonal component in the spin vector, they obtained spacelike spiral trajectories that wind around a reference null geodesics, or equivalently, a reference trajectory for a spinning massless particle with zero orthogonal spin component in the spin vector. They argue that, for ``reasonable cosmologies, redshifts, and atomic periods", the physical distance between the spiral and the null geodesic is of the order of the wavelength, even though according to their analysis it is in principle unbounded.

In their more recent work \cite{Duval2018hzh}, Duval, Marsot, and Schücker extended the analysis to Schwarzschild spacetimes \eqref{eq:schwarzschild}. For the numerical simulations, they assumed initial conditions at the perihelion, the point of closest approach to the star on the trajectory. From their perturbative analysis, they recover two deflection angles, one between the trajectory and the geodesic plane, given by: 

\begin{equation}\label{eq:trajectoryperturbation}
    \beta \sim -\left(1- \frac{2GM}{r_0}\right)\frac{\chi \lambda_0}{2\pi r_0} ,
\end{equation}
and one between the geodesic plane and the momentum carried by the spinning photon:

\begin{equation}\label{eq:momentumperturbation}
    \gamma \sim \chi\frac{GM \lambda_0}{2\pi r_0^2}.
\end{equation}
Here, $\chi=\pm 1$ is the photon helicity. This second deflection angle is proportional to the one presented in \eqref{eq:DeflectionSHE}, derived in \cite{SHE_QM1} from certain approximations applied to field equations, which we discuss further down in section \ref{sec:staticquantum}. It is reassuring that the deflection angle comes out similar with two completely different methods. Despite the previously mentioned shortcomings of the Souriau--Saturnini equations, the authors in \cite{Duval2018hzh}\footnote{Despite this agreement, the workaround for the 'flat space problem' used in \cite{Duval2018hzh}, namely to simply go to the cosmological setting with positive $\Lambda$, to make the problem go away, seems at least on a conceptual level not a very pleasing resolution of the issue.} seem to be able to reproduce these results from \cite{SHE_QM1}. In contrast to \cite{SHE_QM1}, the authors in \cite{Duval2018hzh} provide a clear presentation of their perturbative calculations.

We now give a short discussion about their results and the underlying assumptions. It strikes us as odd that their momentum perturbation \eqref{eq:momentumperturbation}, and the trajectory perturbation \eqref{eq:trajectoryperturbation} come out with a different sign. For the far field asymptotics this seems implausible. Additionally, the trajectory perturbation seems to be independent of the mass, and, between the surface of the Sun and the Earth, it is significantly larger than the momentum perturbation  [$2\beta\sim 10^{-11}{''}$, while $2\gamma\sim 10^{-16}{''}$].

Considering the trajectories in figure \ref{fig:GSHE}, calculated from the equations in section \ref{sec:staticquantum}, that also lead to the same prediction for the deflection angle, one might question whether the assumptions for the perturbative calculations are justified. In figure \ref{fig:GSHE}, one sees that the force, in the direction orthogonal to the geodesic plane, originating from the spin-curvature interaction is not monotone. Therefore, a perturbative expansion around the perihelion might not be justified for light coming from a far away source. For real physical observations it also doesn't seem practical to fix the spin initial conditions at the perihelion of the trajectory. For an actual experiment, this would need to be done either at the location of the emitter or the location of the observer. 

As a final remark to this section, we would like to point out that the situation changes significantly if one considers the Pirani-SSC instead of the Tulczyjew--Dixon SSC. This issue has also been mentioned in a recent paper by Marsot \cite{marsot}. It was shown in \cite{bailyn1977pole} that the Pirani-SSC can be derived from the pole-dipole approximation, if one assumes the stress energy tensor to be traceless ($T^\mu_{\hphantom{\mu}\mu}=0$), on top of the assumption that it be divergence free. Note that both these assumptions are compatible with the stress energy tensor for Maxwell fields. It was argued before \cite{mashhoon1975massless} that, under the Pirani-SSC, massless particles with spin follow ordinary null geodesics, and hence no trace of a G-SHE is present. A similar derivation has been carried out in more detail in \cite{bailyn1977pole,bailyn1981pole}, where it was shown that this is true as long as the assumption $\Vec{p}\cdot \Vec{S}\neq 0$ holds, where $\Vec{p}$ is the spatial part of the momentum, and $\Vec{S}:= (S_1, S_2, S_3)= (S^{23}, S^{31}, S^{12})$. Other aspects of the massless MPD equations with the Pirani-SSC have been discussed by several authors \cite{duval1978conformal, semerak2015spinning,bini2006massless}.

\section{G-SHE from Relativistic Quantum Mechanics} \label{sec:quantum}

The first connection between the motion of spinning particles in curved spacetime and the SHE was introduced by B\'erard and Mohrbach in 2006 \cite{SHE_QM2}. The authors studied the adiabatic evolution of a Dirac particle by using the Foldy--Wouthuysen transformation \cite{FW_original} and presented a generalization of this method for arbitrary spin-fields by using the Bargmann--Wigner equations \cite{BargmannWigner_original}, and a generalized version of the Foldy--Wouthuysen transformation \cite{FW_generalization1,FW_generalization2}. In this way, the position operator of spinning particles acquires an anomalous contribution, related to a non-Abelian Berry connection \cite{SHE_QM2}. Based on this method, Gosselin, B\'erard, and Mohrbach studied the G-SHE of electrons \cite{SHE_Dirac} (similar results were also presented by Silenko et al. \cite{Silenko2005}, albeit without mentioning the term SHE) and photons \cite{SHE_QM1} in a static gravitational field. 

Although not explicitly interested in SHEs, the Foldy--Wouthuysen transformation was also used by Obukhov et al. in order to study the dynamics of Dirac particles in curved spacetime \cite{obukhov2001,obukhov2009,obukhov2011,obukhov2013spin,obukhov2017general}. One important claim discussed in \cite{obukhov2009,obukhov2011,obukhov2013spin,obukhov2017general} is that the linearized MPD equations are obtained as an approximation to the Dirac equation. This will be discussed in more detail in section \ref{sec:diracmpd}

In this section, we will briefly review the G-SHE of photons in a static gravitational field, following the work of Gosselin et al. \cite{SHE_QM1}. We focus on this particular paper because, to our knowledge, it is the only one using techniques adapted from Relativistic Quantum Mechanics in order to describe the propagation of photons in curved spacetime, and the results can easily be compared with the other approaches from sections \ref{sec:spinningparticles} and \ref{sec:maxwell}. The resulting equations of motion are presented and discussed, and connection with the SHE-L in inhomogeneous optical media will be emphasized.

\subsection{Photons in a Static Gravitational Field}\label{sec:staticquantum}

Here we will consider the G-SHE of photons in a static gravitational field, as described by Gosselin, B\'erard and Mohrbach \cite{SHE_QM1}. In this approach, the authors describe electromagnetic waves using the Bargmann--Wigner equations of a massless spin-$1$ field. In general, the Bargmann--Wigner equations describe massive or massless free spin-$j$ fields, and consist of $2j$ coupled partial differential equations, each equation having a similar structure as a Dirac equation \cite{BargmannWigner_original, RelativisticQM}. Considering the case of a spin-$1$ field in the curved spacetime described by the metric $g_{\mu \nu}$, the Bargmann--Wigner equations take the following form:

\begin{align}
(-i \hbar \gamma^\mu \nabla_\mu + m )_{\alpha_1 \alpha^{'}_1} \Psi_{\alpha^{'}_1 \alpha_2} & = 0, \\
(-i \hbar \gamma^\mu \nabla_\mu + m )_{\alpha_2 \alpha^{'}_2} \Psi_{\alpha_1 \alpha^{'}_2} & = 0,
\end{align}
where the field $\Psi_{\alpha_1 \alpha_2}$ is a completely symmetric $4$-spinor of rank $2$, the primed indices are contracted, the gamma matrices satisfy $\{ \gamma^\mu, \gamma^\nu \} = 2 g^{\mu \nu}$, and $\nabla_\mu = \partial_\mu + \omega_\mu$ is the covariant derivative for spinor fields. When setting $m = 0$, it can be shown that these equations are equivalent to the homogeneous Maxwell's equations \cite{RelativisticQM}.

Since now we have a description of electromagnetic waves in terms of coupled Dirac equations, one can import certain methods that are generally used in the Relativistic Quantum Mechanics of electrons. An example of such a method is the Foldy--Wouthuysen transformation \cite{FW_original, RelativisticQM1}, which was originally applied to the massive Dirac equation in order to understand its non-relativistic limit. The Foldy--Wouthuysen transformation consists of a unitary transformation, acting on the basis in which the states and the Dirac Hamiltonian are represented in such a way that the $\alpha$-matrices are eliminated, and the resulting Hamiltonian is in diagonal form. This is always possible for a free Dirac electron, but, in the presence of other external fields (electromagnetic or gravitational), the transformation might only be performed in a perturbative sense, the resulting Hamiltonian being diagonal only to some order in $\hbar$. Generalizations of the Foldy--Wouthuysen transformation to other spin-fields were introduced in \cite{FW_generalization1} for spin-0 and spin-1 fields, and in \cite{FW_generalization2} for arbitrary spin-fields. 

In order to obtain the equations of motion describing the G-SHE of photons in a static gravitational field, Gosselin et al. \cite{SHE_QM1} used a generalized Foldy--Wouthuysen transformation, together with their semiclassical diagonalization procedure described in \cite{SHE_QM2,Gosselin_diagonalization}. Even though their results describe a general static spacetime with torsion \cite{SHE_QM1}, here we will restrict our attention to the particular case of a Schwarzschild background, with the metric expressed in isotropic coordinates, as given in equation \eqref{eq:isotropic}.

In this case, the following equations of motion, describing the G-SHE of photons, were obtained by Gosselin et al. \cite{SHE_QM1}:

\begin{equation}\label{eq:SHESchwarzschild}
\Dot{\mathbf{p}} = - c p \nabla F,
 \qquad  \qquad
\Dot{\mathbf{x}} = c \frac{\mathbf{p}}{p} F + \sigma \frac{\Dot{\mathbf{p}} \times \mathbf{p}}{p^3},
\end{equation}
where $F = \frac{V}{W}$ (see equations \eqref{eq:isotropic} and \eqref{eq:isotropic1} for the definitions of $V$ and $W$) contains the metric components, $\sigma=\pm \hbar$ is the photon helicity, $p = {h}/{\lambda}$ is the magnitude of the photon momentum, and the vector notation is $\mathbf{p} = (p_x, p_y, p_z)$, $\mathbf{x} = (x, y, z)$. 

There is a small difference between the equations of motion presented here in equations \eqref{eq:SHESchwarzschild}, and the equations of motion from \cite[eq.(24)]{SHE_QM1}, where the authors used a wrong formula for $F$ in the last step of their calculation. While this does not seem to affect the final results in a drastic way, the error propagated into other papers as well \cite{SHE_pictures}.  

The G-SHE is given by the second term in the expression of $\Dot{\mathbf{x}}$. Clearly, this is a helicity dependent correction, which vanishes when we neglect the helicity of the photon. In this case, the equations of motion reduce to the usual null geodesics, and describe ordinary light bending around a Schwarzschild black hole. Also, the G-SHE correction term is proportional to the wavelength $\lambda$ of the photon, since $\Dot{\mathbf{p}} \propto p \propto \lambda^{-1}$, $\mathbf{p} \propto \lambda^{-1}$, and $p^3 \propto \lambda^{-3}$. Thus, the G-SHE vanishes in the limit of very short wavelengths or infinitely high frequencies.

\subsection{Predictions of the theory} \label{sec:quantumpredictions}

It is important to notice the direction of the G-SHE correction term in equations \eqref{eq:SHESchwarzschild}. Given the spherical symmetry of the Schwarzschild black hole, $\Dot{\mathbf{p}}$ is directed along the gradient of the gravitational field, and $\mathbf{p}$ will correspond to the direction of propagation of the photon. Thus, the G-SHE correction term will be perpendicular to both the gradient of the gravitational field and the direction of propagation of the photon. In the case of radial trajectories, the G-SHE will vanish. 

For a Schwarzschild black hole, trapped null geodesics occur only at a fixed radial location, given by $r = \frac{3G M}{c^2}$. These null geodesics constitute the photon sphere, and determine the shadow of the black hole (see e.g. \cite{Shadow_deg1} for a discussion of black hole shadows). One can consider a notion of effective photon sphere in the context of the G-SHE. By effective photon sphere, we now mean trapped curves associated with the equations of motion \eqref{eq:SHESchwarzschild}. Considering any photon initially located on the photon sphere, the G-SHE correction term will be different from zero. However, since $\Dot{\mathbf{p}}$ is directed along a radial direction, and $\mathbf{p}$ is tangential to the photon sphere, their cross product will also be tangential to the photon sphere. Even though the G-SHE will deflect the photon from its original trajectory, the photon will not leave the photon sphere. This contradicts the findings in \cite{wang2016effect}, where it was predicted that the gravitational spin-orbit coupling induces a helicity-dependent splitting of the photon sphere into two effective photon spheres of different radii. Another argument against the splitting of the photon sphere is that both the Schwarzschild spacetime and Maxwell's equations are parity invariant, while a splitting of the photon sphere would violate this basic symmetry principle.

Based on the equations of motion \eqref{eq:SHESchwarzschild}, Gosselin et al. proposed the following correction to the ordinary light bending deflection angle \cite{SHE_QM1}:

\begin{equation}\label{eq:Deflection}
\Delta \phi = \frac{4 G M}{c^2 r_0} \left(1 - \frac{\sigma}{\hbar} \frac{\lambda}{2\pi r_0} \right),
\end{equation}
where $r_0$ represents the distance of closest approach between the light ray and the center of the gravitational source. The first term in this equation represents the ordinary light bending deflection angle, well known from General Relativity, while the second term represents the polarization-dependent G-SHE correction. However, equation \eqref{eq:Deflection} cannot be correct. The additional deflection due to the G-SHE lies in a plane orthogonal to the ordinary light bending plane, thus, one should treat them separately. This statement is justified by the equations of motion \eqref{eq:SHESchwarzschild}, where the G-SHE correction term is proportional to $\mathbf{x} \times \mathbf{p}$ (this is because $\Dot{\mathbf{p}} \propto \nabla F \propto \mathbf{x}$). Looking in the plane orthogonal to the ordinary light bending plane, the deflection angle coming from the G-SHE should be:

\begin{equation}\label{eq:DeflectionSHE}
\Delta \phi_{SHE} = \frac{\sigma}{\hbar} \frac{4 G M}{c^2 r_0}  \frac{\lambda}{2\pi r_0} = \chi \frac{2 G M \lambda}{ \pi c^2 r_0^2},
\end{equation}
where $\chi = \pm 1$. This last formula is proportional to the deflection angle predicted in \cite{Duval2018hzh}, and discussed here in equation \eqref{eq:momentumperturbation}. Also, the classical nature of the G-SHE is emphasized here, since the deflection angle does not depend on $\hbar$. 

One of the main disadvantage of the method used in \cite{SHE_QM1} is that the use of the Bargmann--Wigner equations blurs the connection with Maxwell's equations, while the Foldy--Wouthuysen transformation, and the semiclassical diagonalization procedures unnecessarily introduce Planck's constant, giving the general impression that the G-SHE is of Quantum Mechanical origin. In section \ref{sec:SHEL_eq}, the SHE-L was directly derived from Maxwell's equations, without the need of using any Quantum Mechanical notions. Similar arguments should apply for the case of Maxwell's equations on a curved background. Another drawback of the approach of Gosselin et al. is that their treatment is limited to static spacetimes, and it is not clear how the method should be extended to more general spacetimes. 

An alternative derivation of equations \eqref{eq:SHESchwarzschild} can be obtained by treating the Schwarzschild spacetime as an effective inhomogeneous medium. By using the equivalence between Maxwell's equations in curved spacetime and inside a material in flat spacetime, as discussed in section \ref{sec:equivalence}, an effective refractive index can be attributed to the Schwarzschild spacetime, $n = 1/F$, and the same methods as for the SHE-L in inhomogeneous optical media can be applied. For example, equations \eqref{eq:SHESchwarzschild} can be easily obtained by inserting $n = 1/F$ into equations \eqref{eq:SHE_Ruiz}.

\begin{figure}[t!]
\centering
\centering
  \subfloat[\label{fig:ss3}]{%
    \includegraphics[width=.40\textwidth]{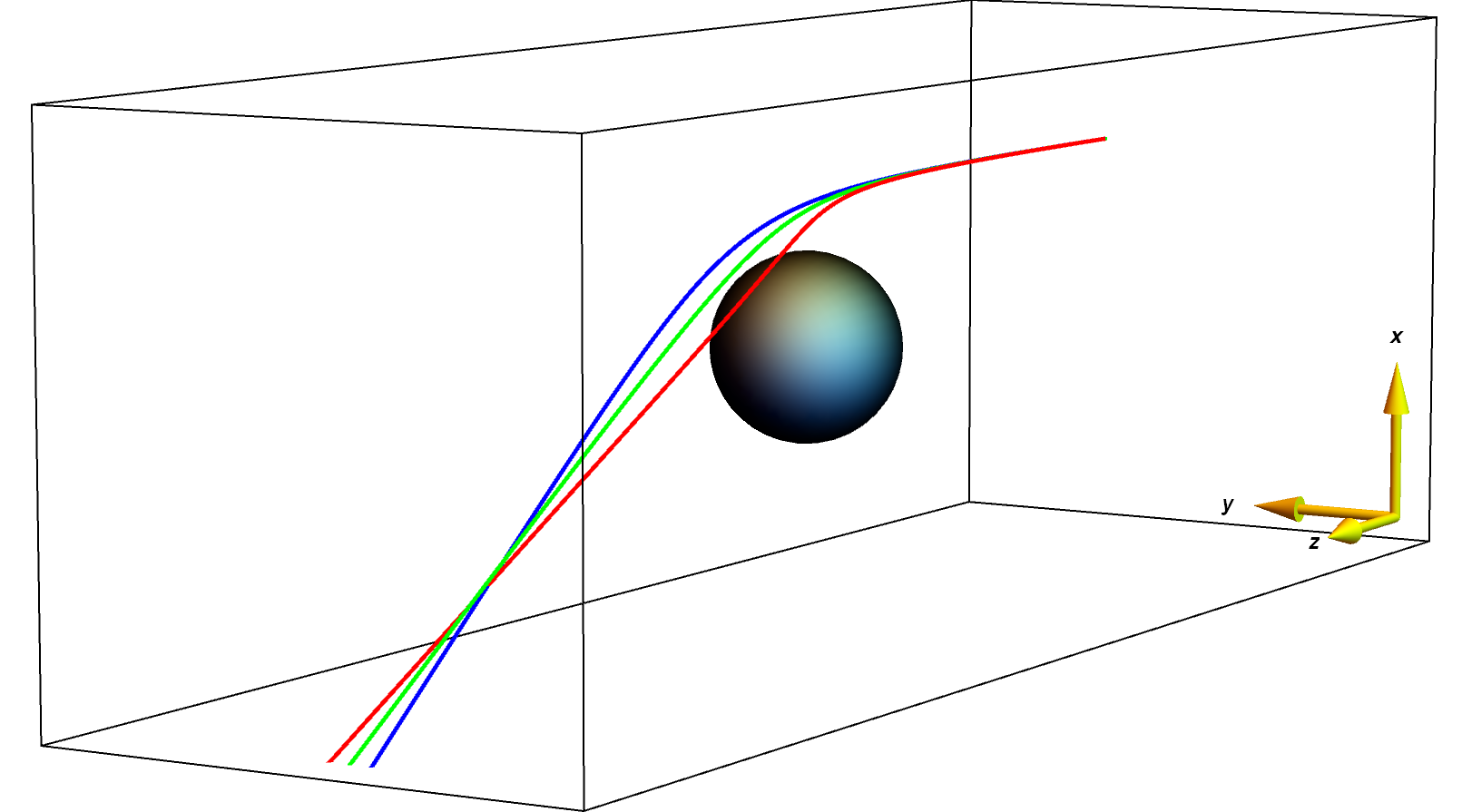}}\hfill
  \subfloat[\label{fig:ss2}]{%
    \includegraphics[width=.27\textwidth]{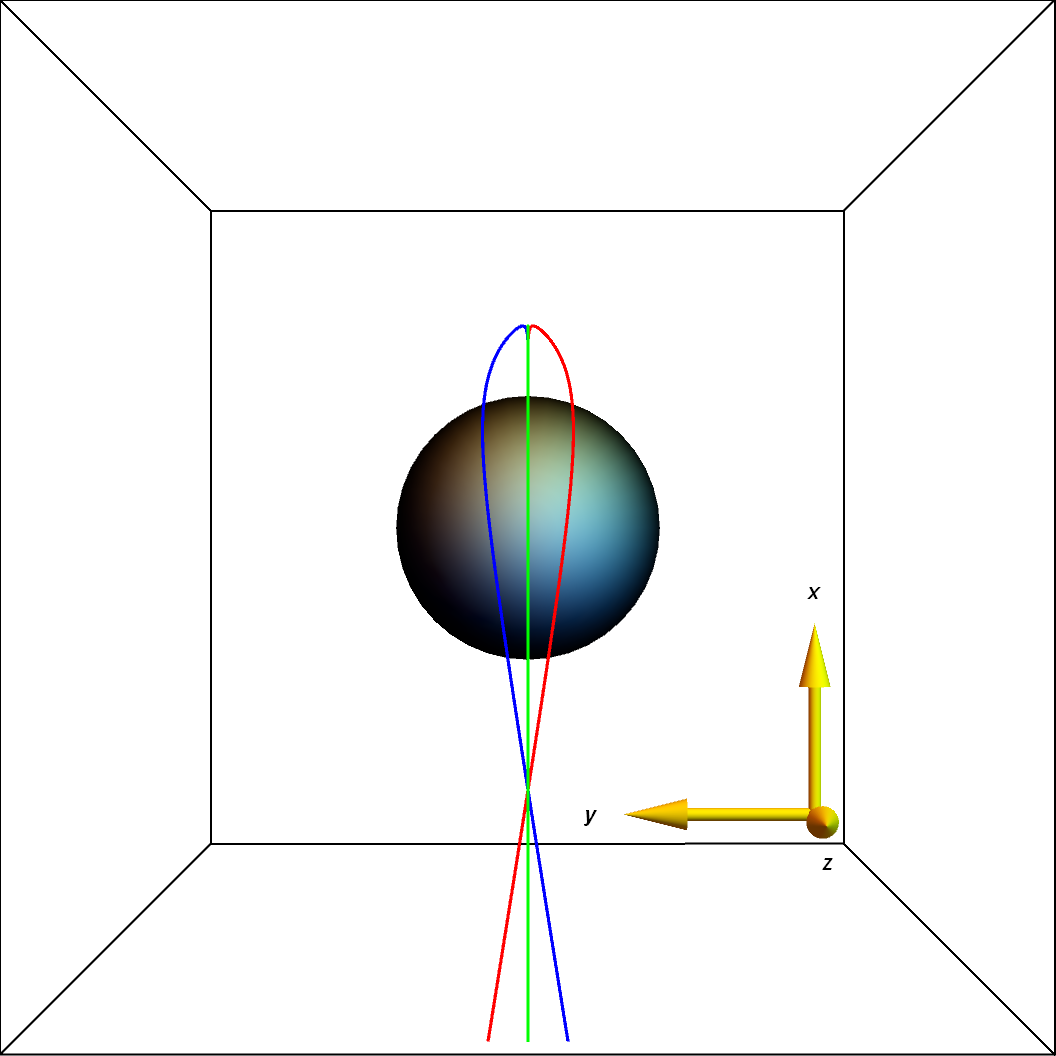}}\hfill
   \subfloat[\label{fig:ss1}]{%
    \includegraphics[width=.20\textwidth]{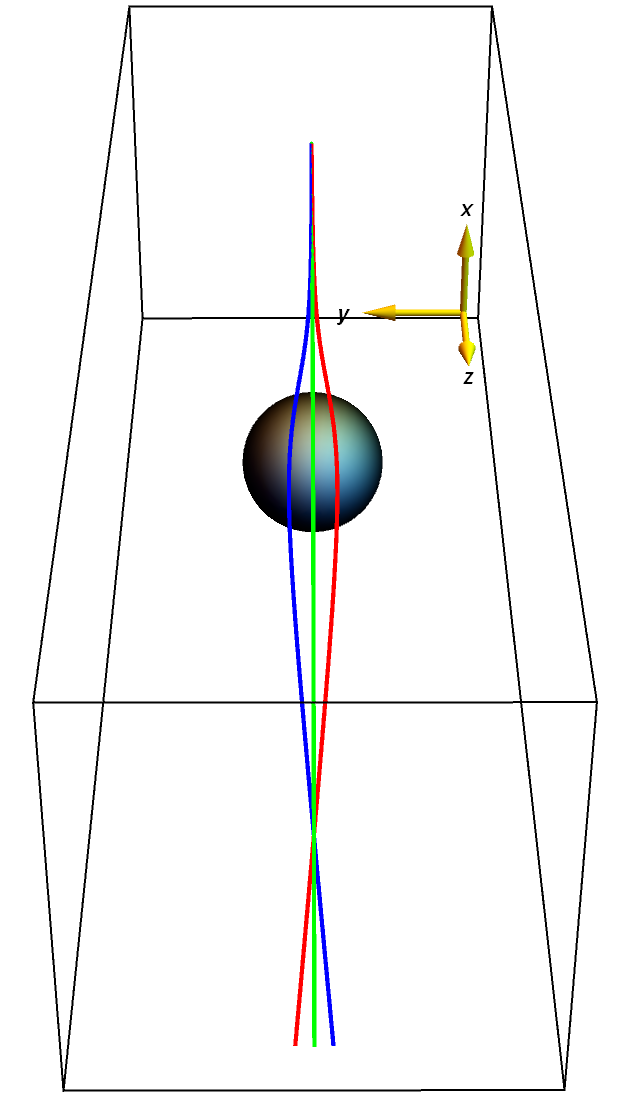}}\hfill
\caption{G-SHE of photons around a Schwarzschild black hole. Different viewing angles for the same trajectories are represented. The magnitude of the effect is amplified for visualization purposes. The blue and the red trajectories correspond to photons having opposite helicities, $\sigma= \pm \hbar$, while the green trajectory represents a null geodesic ($\sigma = 0$), undergoing only ordinary light bending.}
\label{fig:GSHE}
\end{figure}

As discussed in \cite{SOI_review}, all spin-orbit interactions of light can be understood in terms of couplings between the different forms of angular momentum that light can carry. In the case considered by Gosselin et al. \cite{SHE_QM1}, photons only have SAM and EOAM with respect to the black hole, thus the only possible spin-orbit coupling is between these two forms of angular momentum. Given the spherical symmetry of the Schwarzschild spacetime, one can show that equations \eqref{eq:SHESchwarzschild} possess the following integral of motion:

\begin{equation}
    \mathbf{J} = \mathbf{x} \cross \mathbf{p} + \sigma \frac{\mathbf{p}}{p}, \qquad \qquad \dot{\mathbf{J}} = 0. 
\end{equation}

The same holds true for the SHE-L in spherically-symmetric inhomogeneous optical media \cite{SOI_review, SHE_original}, and this emphasizes the direct connection between the conservation of the total angular momentum of light and the helicity-dependent corrections to the equations of motion.

In order to provide some intuition about how the G-SHE affects the propagation of light around a Schwarzschild black hole, we numerically integrated equations \eqref{eq:SHESchwarzschild}. An example is presented in figure \ref{fig:GSHE}, where we start at a common point with three different trajectories. The only difference in the initial conditions is the helicity.  One can see that the G-SHE results in a helicity-dependent transverse shift of the trajectories, and the motion is no longer restricted to a plane, as in the case of the null geodesic. The Schwarzschild black hole acts as a Stern--Gerlach magnet for photons of opposite helicity. Other examples of numerically integrated G-SHE trajectories can also be found in \cite{SHE_pictures}.

\section{G-SHE from Geometrical Optics} \label{sec:maxwell}

The standard treatment for the propagation of electromagnetic waves in General Relativity is achieved by investigating Maxwell's equations in curved spacetime. Null geodesics can be obtained from Maxwell's equations by considering the lowest order geometrical optics approximation \cite{MTW,Perlick2000,Perlick2004}. However, as we saw in the previous sections, at this level of the approximation, there is no influence of the polarization degree of freedom on the trajectories. In order to obtain a theoretical description of the G-SHE, higher order terms should be considered in the geometrical optics approximation.

Starting with Maxwell's equations in curved spacetime, and by considering certain corrections to the standard geometrical optics approximation, several authors obtained polarization-dependent trajectories for light rays in a curved spacetime \cite{Frolov,Frolov2,covariantSpinoptics,spinorSpinoptics,spinorSpinoptics2} (see also \cite{Harte2018} for a more general discussion). However, some of the predictions presented in these papers are in contradiction with the results discussed in sections \ref{sec:spinningparticles} and \ref{sec:quantum}. For example, polarization-dependent trajectories were predicted in \cite{Frolov,covariantSpinoptics}, on a Kerr spacetime. However, this effect disappears in the limit of a Schwarzschild spacetimes, in contrast to what we discussed in the previous sections. 

Here, we will review the main features of these approaches, focusing in particular on \cite{Frolov,covariantSpinoptics}. We start by reviewing the standard geometrical optics approximation for Maxwell's equations in curved spacetime. In the lowest order expansion, this leads to the well-known results that light rays follow null geodesics, and the polarization vector is parallel-transported along the null geodesic, leading to the gravitational Faraday rotation of the polarization vector. The gravitational Faraday rotation represents the starting point for the modified geometrical optics proposal presented in \cite{Frolov,Frolov2,covariantSpinoptics,spinorSpinoptics,spinorSpinoptics2}, where a modified eikonal ansatz was proposed.  

\subsection{Geometrical Optics and Gravitational Faraday Rotation} \label{sec:Frolov_GO}

In this section we will review the main features of the geometrical optics approximation for Maxwell's equations in curved spacetime, which is well known in General Relativity \cite{MTW}. The general argument is very similar to what we already discussed in section \ref{sec:SHEL_eq}, but there are a some key differences, which will be emphasized along the way. It is important to compare these two approaches in detail, since there is some disagreement between their predictions for the G-SHE, as we will see in what follows.  

We begin by considering a stationary spacetime described by some metric tensor $g_{\mu \nu}$. The propagation of electromagnetic waves within a given spacetime can be described by the vector potential $A_\mu$, satisfying the Lorentz gauge condition

\begin{equation}
    \nabla_\mu A^\mu = 0,
\end{equation}
and the wave equation

\begin{equation}
    \nabla^\nu \nabla_\nu A_\mu - R\indices{_{\mu}^{\nu}} A_\nu = 0,
\end{equation}
where $\nabla_\mu$ is the covariant derivative, and $R_{\mu \nu}$ is the Ricci tensor.

The central assumption of the geometrical optics approximation is that the wavelength of light, $\lambda=2 \pi \omega^{-1}$, is much smaller than any other characteristic length scale $L$ of the problem. When considering the propagation of light through a medium, this length scale is given by the distance over which the parameters of the medium change significantly, and in the case of light propagating on a curved spacetime, the length scale is given by the variation scale of the spacetime curvature. Under this assumption, it is expected that the vector potential $A_\mu$ can be split into a slowly varying complex amplitude $a_\mu$ and a fast oscillating real phase $S$:

\begin{equation} \label{eq:eikonal-ansatz}
    A_\mu = a_\mu e^{i S / \epsilon},
\end{equation}
where $\epsilon$ is a dimensionless expansion parameter.

This is generally called the eikonal ansatz. The same results are obtained if one uses the eikonal ansatz to expand other quantities, such as the Faraday tensor \cite{spinorSpinoptics,spinorSpinoptics2} or the Riemann--Silberstein vector \cite{Frolov}. The geometrical optics equations are obtained by inserting the eikonal ansatz into the Lorentz gauge condition and into the wave equation. Afterwards, the results are examined order by order in the expansion parameter, $\epsilon$. From the Lorentz gauge condition, at order $\epsilon^{-1}$, we obtain:

\begin{equation}\label{eq:GO1}
    a^\mu k_\mu = 0,
\end{equation}
where we have defined $k_\mu = \nabla_\mu S$. Thus, the amplitude vector $a_\mu$ is orthogonal to the wave vector $k_\mu$. The wave equation at order $\epsilon^{-2}$ gives:

\begin{equation}\label{eq:GO2}
    k^\mu k_\mu = 0.
\end{equation}
This means that the gradient of the phase is null. Also, since $k_\mu$ is a gradient, it follows that it will satisfy the null geodesic equation:

\begin{equation}
    k^\nu \nabla_\nu k^\mu = 0.
\end{equation}

In this way, the classical result that light rays follow null geodesics is recovered from the wave equation. It is one of the main results of the geometrical optics approximation for Maxwell's equations in curved spacetime.

By examining the wave equation at order $\epsilon^{-1}$, we obtain the following transport equation:

\begin{equation}\label{eq:GO3}
    k^\nu \nabla_\nu a^\mu + \frac{1}{2} a^\mu \nabla_\nu k^\nu = 0.
\end{equation}

Following \cite{covariantSpinoptics}, we can split the amplitude $a_\mu$ into a real amplitude $a$ and a complex unit vector $l^\mu$, which will describe the polarization degree of freedom:

\begin{equation}\label{eq:polarizationansatz}
    a_\mu = a l_\mu, \qquad \qquad l^\mu \Bar{l}_\mu = 1. 
\end{equation}

Substituting equation \eqref{eq:polarizationansatz} into equation \eqref{eq:GO1}, we see that the polarization vector $l^\mu$ is orthogonal to the wave vector $k_\mu$:

\begin{equation}
    k_\mu l^\mu = 0.
\end{equation}

By substituting equation \eqref{eq:polarizationansatz} into equation \eqref{eq:GO3}, we obtain:

\begin{align}
    \nabla_\mu (a^2 k^\mu) & = 0, \label{eq:photoncons} \\
    k^\nu \nabla_\nu l^\mu & = 0. \label{eq:polarizationpropagation}
\end{align}

Equation \eqref{eq:photoncons} can be interpreted as the conservation of the photon number along the null geodesic, and equation \eqref{eq:polarizationpropagation} represents the parallel-propagation equation for the polarization vector along the null geodesics. This is the second important result of the geometrical optics approximation for Maxwell's equations in curved spacetime.

Following \cite{covariantSpinoptics}, we can introduce an orthonormal basis $(u^\mu, n^\mu, e_1^\mu, e_2^\mu)$. Since the considered spacetime is stationary, we have the following stationary Killing vector field:

\begin{equation}
    \xi = \frac{\partial}{\partial t},
\end{equation}
and we introduce the notation $h = -\xi^\mu \xi_\mu$. Now, we can define $u^\mu$ and $n^\mu$ as in \cite{covariantSpinoptics}:

\begin{align} \label{eq1}
u^\mu & = \frac{\xi^\mu}{\sqrt{h}}, \\
n^\mu & = \frac{\sqrt{h}}{\omega} k^\mu - u^\mu,
\end{align}
where $\omega = -\xi^\mu k_\mu $ is the frequency measured by a stationary observer. The other two spacelike unit vector fields, $e_1^\mu$ and $e_2^\mu$, can be obtained by Fermi--Walker transport along the vector $n^\mu$ \cite{Frolov, covariantSpinoptics}. In this case, the vector fields $e_1^\mu$ and $e_2^\mu$ will represent a linear polarization basis, and we can define the circular polarization basis in the following way:

\begin{equation} \label{eq:polarizationansatz1}
    m^\mu = \frac{1}{\sqrt{2}} (e_1^\mu + i \sigma e_2^\mu),
\end{equation}
where $\sigma = +1$ corresponds to right-handed circular polarization, and  $\sigma = -1$ corresponds to left-handed circular polarization. We will refer to $\sigma$ as the helicity. Using this circular polarization base vector field $m^\mu$, we can write the polarization vector $l^\mu$ as in \cite{covariantSpinoptics}:

\begin{equation} \label{eq:polarizationansatz2}
    l^\mu = m^\mu e^{i \phi},
\end{equation}
where $\phi=\phi(x)$ is a real function of the spacetime coordinates, and the amplitude $a^\mu$ will take the following form:

\begin{equation} \label{eq:polarizationansatz3}
    a^\mu = a l^\mu =  a (e_1^\mu + i \sigma e_2^\mu)  e^{i \phi}.
\end{equation}

This form of the amplitude is very similar to the one considered in section \ref{sec:SHEL_eq}, with the small difference that here only one circular polarization mode is considered to be active (we have either a right-handed or a left-handed circular polarization mode, depending on the helicity $\sigma$). This should not affect the following calculations, since the polarization dynamics should be decoupled in a circular polarization basis. However, the method presented here, and the method discussed in section \ref{sec:SHEL_eq} are quite different, since in section \ref{sec:SHEL_eq} the field Lagrangian was projected onto the circular polarization eigenmodes, while here there is no such projection, and the transport equation is already given. 

Recalling that $l^\mu$ is parallel-transported along the null geodesic generated by $k^\mu$, equation \eqref{eq:polarizationpropagation} becomes:

\begin{equation}
     k^\nu \nabla_\nu l^\mu = k^\nu \nabla_\nu (m^\mu e^{i \phi}) = 0.
\end{equation}

Following the derivation in \cite{covariantSpinoptics}, we can write the propagation equation for the phase $\phi$ along null geodesics generated by $k^\mu$:

\begin{equation}
    k^\nu \nabla_\nu \phi = \frac{1}{2} \sigma u_\rho k_\lambda \varepsilon^{\mu \nu \rho \lambda} \nabla_\nu u_\mu,
\end{equation}
where $\varepsilon^{\mu \nu \rho \lambda}$ is the Levi--Civita tensor. This equation describes the evolution of the phase function $\phi$ along the null geodesic generated by $k^\mu$. This effect is known as the gravitational Faraday rotation \cite{FaradayRotation1,FaradayRotation2,FaradayRotation3,FaradayRotation4,FaradayRotation5,FaradayRotation6,Schneiter2018}.

At least in the case considered here, the extra phase variation arising as a consequence of the gravitational Faraday rotation is a phenomenon strictly related to the non-static nature of the spacetime. More explicitly, the extra phase variation is proportional to the $g_{0 i}$ off-diagonal terms in the metric \cite{Frolov,covariantSpinoptics} (this is clearly presented in equation (102) from \cite{Frolov}). If we consider a Kerr spacetime in Boyer-–Lindquist coordinates, with spin parameter $a$, then the variation of $\phi$ along a null geodesic generated by $k_\mu$ would be proportional to $a$.

\subsection{Modified Geometrical Optics}\label{sec:modgo}

The standard geometrical optics approximation predicts the gravitational Faraday rotation, and the trajectories of light rays are null geodesics, independent of the polarization. In order to take into account the influence of the polarization on the null geodesics, a modified geometrical optics procedure (also called ``spinoptics" by some authors) was presented, first by Frolov and Shoom \cite{Frolov}, and later on by Yoo \cite{covariantSpinoptics} and Dolan \cite{spinorSpinoptics,spinorSpinoptics2}. The main idea is that the additional phase factor coming from the gravitational Faraday rotation should be interpreted as a correction term to the original eikonal ansatz, considered in equation \eqref{eq:eikonal-ansatz}. By adopting this approach, the eikonal ansatz is modified in the following way:

\begin{align}
    S \rightarrow \Tilde{S} & = S + \phi, \\
    A_\mu & = \Tilde{a}_\mu e^{i \Tilde{S} / \epsilon}. 
\end{align}

This new eikonal ansatz looks somewhat similar to what Bliokh et al. considered in \cite{Bliokh2004}, where an extra Berry phase was included in the eikonal ansatz. However, at this point, it is not clear if we can identify the gravitational Faraday rotation with the Berry phase of electromagnetic waves propagating in curved spacetime. The main reason behind this is the fact that the gravitational Faraday rotation, as presented in \cite{Frolov,covariantSpinoptics}, vanishes in static spacetimes, such as the Schwarzschild spacetime. On the other hand, from the results of Gosselin et al. \cite{SHE_QM1} we clearly see that the Berry phase is non-vanishing and plays a key role for the G-SHE of photons propagating in static spacetimes. 

Following the same steps as for the standard geometrical optics approximation, the following equations are obtained \cite{covariantSpinoptics}:

\begin{align}
    \Tilde{a}^\mu q_\mu & = 0, \\ 
    q^\mu q_\mu & = 0, \\
    \nabla_\mu (\Tilde{a}^2 q^\mu) & = 0, \\
    q^\mu \nabla_\mu \Tilde{\phi} & = 0,
\end{align}
where $q_\mu = \nabla_\mu \Tilde{S} -\sigma \phi_\mu$, $\phi_\mu = \frac{1}{2} \varepsilon_{\mu \nu \rho \lambda} u^\nu \nabla^\rho u^\lambda$, and $\tilde{a}^\mu = a \tilde{m}^\mu e^{i \tilde{\phi}}$. These equations indicate that the trajectories generated by $q^\mu$ are null, the photon number is conserved, and the phase $\Tilde{\phi}$ is constant along the null trajectories generated by $q^\mu$.

However, since the modified wave vector $q_\mu$ is no longer a gradient, the following equation of motion is obtained for light rays in this modified geometrical optics approximation \cite{covariantSpinoptics}:

\begin{equation} \label{eq:modgo_force}
    q^\nu \nabla_\nu q^\mu = \sigma f\indices{^\mu_\nu} q^\nu ,
\end{equation}
where $f_{\mu \nu} = \nabla_\mu \phi_\nu - \nabla_\nu \phi_\mu$. 

This equation is similar to that of the motion of charged particles under the influence of an electromagnetic field. Here, the role of the charge is played by the polarization $\sigma$, and the role of the electromagnetic vector potential is played by $\phi_\mu$.

The results of this modified geometrical optics approach can also be obtained by considering an effective metric, as shown by Frolov et al. \cite{Frolov}. Considering the case of a Kerr spacetime in Boyer--Lindquist coordinates, with $a$ representing the black hole spin parameter, the effective metric with modified geometrical optics corrections can be written as:

\begin{equation}
    g_{\mu \nu} = \begin{pmatrix} 
g_{t t} &  g_{t r} &  g_{t \theta}  & g_{t \phi} \\
g_{r t} & g_{r r} & 0 & 0 \\ 
g_{\theta t} & 0 & g_{\theta \theta} & 0 \\
g_{\phi t} & 0 & 0 & g_{\phi \phi}
\end{pmatrix} ,
 \end{equation}
 where the effective correction terms are:
\begin{align}
    g_{r t} = g_{t r} & = \frac{\sigma a}{\omega} f_1 (a, M, r, \theta), \\
    g_{\theta t} = g_{t \theta} & = \frac{\sigma a}{\omega} f_2 (a, M, r, \theta).
\end{align}

The explicit form of the functions $f_1$ and $f_2$ can be obtained from \cite[eq. (126)]{Frolov}. The key aspect is that the effective correction terms are proportional to $a$, and vanish when $a = 0$. Thus, there is no effect in the case of a Schwarzschild spacetime, in contrast to what we discussed in sections \ref{sec:ss} and \ref{sec:staticquantum}. Also, the effective correction terms go to zero when we neglect the polarization degree of freedom, and in the limit of high frequencies, similarly as for the SHE-L in section \ref{sec:SHEL_eq}, or for the G-SHE from section \ref{sec:staticquantum}. 

The same modified geometrical optics procedure was applied in \cite{covariantSpinoptics} to study the propagation of gravitational waves, with similar results as presented above. The only difference comes from the fact that gravitational waves are described by a massless spin-2 field, so we have helicity $\sigma = \pm 2$. These claims are in contradiction with the results of Yamamoto \cite{SHE_GW}, which predicted a G-SHE for gravitational waves in Schwarzschild spacetimes.

\section{Linking the Models} \label{sec:MPDequivalence}

In the previous sections we saw that several authors obtained various forms of G-SHEs for massless particles, using completely different methods, and sometimes even resulting in predictions that do not agree. As a starting point towards a deeper understanding of the G-SHE, one could try to show that some connections exist between these apparently different methods. At least for the case of the SouSa equations and the method presented in section \ref{sec:quantum}, such a connection could be expected, since the predicted deflection angles seem to agree in Schwarzschild spacetimes. Unfortunately, no such connections have been explored in the literature.    

However it is reassuring to see that, at least for the case of massive particles, there exists some work linking the approach in section \ref{sec:quantum}, as well as the geometrical optics approach, to linearized MPDT equations. These will be presented in the following. Our hope is that future developments of the G-SHE for massless particles could benefit from this discussion.

\subsection{MPD -- Dirac Equivalence from the Quantum Perspective}\label{sec:diracmpd}

Here we present a sketch of the derivation of the linearized MPDT equations, starting from the massive Dirac equation, as proposed by Obukhov, Silenko, and Teryaev \cite{obukhov2013spin,obukhov2017general}. The central element of their derivation is the application of the Foldy--Wouthuysen transformation, so in this sense it is somewhat similar to the approach presented in section \ref{sec:quantum}. The authors make use of the following representation of a generic metric:

\begin{equation}
    ds^2= V^2dt^2-\delta_{{\hat{a}}{\hat{b}}}W^{\hat{a}}_{\hphantom{\hat{a}}c}W^{\hat{b}}_{\hphantom{\hat{b}}d}(dx^c-K^cdt)(dx^d-K^ddt),
\end{equation}
where $t$ stands for a time coordinate, and $x^a$, with $(a=1,2,3)$, denote local spatial coordinates. They choose the following tetrad, which satisfies the Schwinger gauge $e^{\hat{0}}_a=0$:

\begin{equation}
    e^{\hat{0}}_i=V \delta^{\hat{0}}_i, \qquad e^{\hat{a}}_i= W^{\hat{a}}_{\hphantom{\hat{a}}b}(\delta^b_i-K^b \delta^0_i).
\end{equation}

A particle moves along a worldline $x^\mu(\tau)$, where $(\mu=0,1,2,3)$, and $\tau$ is the proper time. The four-velocity is then $u^\mu=dx^\mu/d\tau$, and $u^\alpha=e^\alpha_\mu u^\mu$, with $(\alpha=0,1,2,3)$ in tetrad components. They use the representation $u^\alpha=(\gamma,\gamma v^{\hat{a}})$, where $\gamma^{-1}=\sqrt{1-v^2}$, and $v^{\hat{a}}$ are the three spatial components of the velocity. As a consequence, we have:

\begin{align}
    u^0&=\frac{dt}{d\tau}=\frac{\gamma}{V} ,\\
    u^a&=\frac{dx^a}{d\tau}=\frac{\gamma}{V}(K^a+ F^a_{\hphantom{a}b}v^b) ,
\end{align}
where:

\begin{equation}
    F^a_{\hphantom{a}b}=V W^{a}_{\hphantom{a}b}.
\end{equation}

The authors start with the Dirac equation:

\begin{equation}\label{eq:dirac}
    \gamma^k\nabla_k \Psi - \frac{\mu}{\hbar}\Psi =0,
\end{equation}
where $\Psi$ is a $4$-spinor, and, upon fixing a tetrad, $\gamma^k$ are the gamma matrices. Equation \eqref{eq:dirac} can be derived from the action $S=\int d^4x \sqrt{-g} \mathcal{L}$, with the Lagrangian density:

\begin{equation}
    \mathcal{L}=\frac{i \hbar}{2}(\overline{\Psi}\gamma^\alpha\nabla_\alpha \Psi -\Psi\gamma^\alpha\nabla_\alpha \overline{\Psi}) -\mu \overline{\Psi} \Psi.
\end{equation}

To obtain a ``Hermitian Hamiltonian"\footnote{In neither paper \cite{obukhov2013spin,obukhov2017general}, did the authors mention which scalar product they are considering.} when writing the Dirac equation in Schr\"{o}dinger form, $i\hbar \frac{\partial \psi}{\partial t}=\mathcal{H}\psi$, the authors introduce the following rescaled wavefunction:

\begin{equation}
    \psi=(\sqrt{-g}e^0_{\hat{0}})^{1/2}\Psi.
\end{equation}

Then, the Hamiltonian is given by:

\begin{equation}
    \mathcal{H}=\gamma^0 m V+ \frac{1}{2}\left(\pi_a F^a_{\hphantom{a}b}\alpha^b+  \alpha_aF^a_{\hphantom{a}b}\pi^a + K^a\pi_a + \pi_ a K^a + \frac{\hbar}{2}(\Xi_a\Sigma^a - \Upsilon \gamma_5)\right),
\end{equation}
where $\pi_a= -i\hbar \partial_a=p_a$\footnote{Note that this choice for the momentum operator is not Hermitian in curved spacetime. If one is only interested in the weak field approximation, this should not be a concern. However, the notion of Hermiticity is dependent on the scalar product, which is not specified in the papers discussed here. }, $\alpha^a= \gamma^0 \gamma^a$, $\Sigma^1=i\gamma^2 \gamma^3$, $\Sigma^2=i\gamma^3\gamma 1$, and $\Sigma^3=i \gamma^1\gamma^2$. Furthermore, one can introduce a pseudoscalar, $\Upsilon$, and a three-vector, $\Xi$:

\begin{equation}
    \Upsilon= V \epsilon ^{abc}\Gamma_{abc}, \qquad \Xi_a= V \epsilon_{abc}\Gamma_0 ^{\hphantom{0}bc}.
\end{equation}

With the methods developed in \cite{silenko2008foldy}, the authors of \cite{obukhov2013spin,obukhov2017general} then proceed to derive the Hamiltonian in the Foldy--Wouthuysen representation \cite{FW_original}. In this step, they linearize in $\hbar$, hence only keeping contributions to the Hamiltonian that are of the zeroth or first order in $\hbar$. The Hamiltonian is decomposed into pieces that commute and anticommute with $\gamma^0$:

\begin{equation}
    \mathcal{H}=\gamma^0 \mathcal{M} + \mathcal{E}+ \mathcal{O}, \qquad \gamma^0\mathcal{M}= \mathcal{M}\gamma^0, \qquad \gamma^0\mathcal{E}= \mathcal{E}\gamma^0, \qquad \gamma^0\mathcal{O}= -\mathcal{O}\gamma^0.
\end{equation}

Hence, the operators $\mathcal{M}$, $\mathcal{E}$ are even, and $\mathcal{O}$ is odd. The Foldy--Wouthuysen representation is given by:

\begin{equation}
    \psi_{FW}=U\psi, \qquad \mathcal{H}_{FW}= U \mathcal{H}U ^{-1} -i \hbar U\partial_t U ^{-1}.
\end{equation}

Assuming the notation $\epsilon= \sqrt{\mathcal{M}^2+\mathcal{O}^2}$, the operator $U$ is given by:

\begin{equation}
    U=\frac{\gamma^0 \epsilon + \gamma^0 \mathcal{M}-\mathcal{O}}{\sqrt{(\gamma^0 \epsilon + \gamma^0 \mathcal{M}-\mathcal{O})^2}}\gamma^0 ,
\end{equation}
and it is unitary ($U^{-1}=U^\dagger$) if $\mathcal{H}=\mathcal{H}^\dagger$. 

After introducing the polarization operator $\Pi=\gamma^0 \Sigma$, the authors of \cite{obukhov2013spin,obukhov2017general} calculate: 

\begin{equation}
    \frac{d\Pi}{dt}=\frac{i}{\hbar}[\mathcal{H}_{FW},\Pi]=\Omega_{(1)}\cross\Sigma + \Omega_{(2)}\cross \Pi,
\end{equation}
where the three-vectors $\Omega_{(1)}$ and $\Omega_{(2)}$ are the operators of the angular velocity of spin precession. Then, the authors of \cite{obukhov2013spin,obukhov2017general} obtained the semiclassical equations describing the motion of the average spin vector $\mathbf{s}$ by evaluating all anticommutators, and omitting powers of $\hbar$ higher than one:

\begin{equation}\label{eq:spinequation}
    \frac{d\mathbf{s}}{dt}=\mathbf{\Omega}\cross \mathbf{s}= (\Omega_{(1)}+ \Omega_{(2)})\cross \mathbf{s}.
\end{equation}
Here, $\{A\cross B\}_a= \epsilon_{abc}A^b B^c$ is the usual vector product in three dimensions. Substituting the semiclassical limit back into the Foldy--Wouthuysen Hamiltonian, they get:

\begin{equation}
    \mathcal{H}_{FW}=\gamma^0 \gamma m V+ \frac{1}{2}\left(K^a p_a + p_ a K^a\right)+ \frac{\hbar}{2}\left(\Pi \cdot \Omega_{(1)}+ \Sigma \cdot \Omega_{(2)}\right).
\end{equation}

From this, the velocity operator is obtained in the following form:

\begin{equation}\label{eq:velocityequation}
   u^a= \frac{dx^a}{dt}= \frac{i}{\hbar}[\mathcal{H}_{FW}, x^a]=\gamma^0 \frac{\partial \gamma m V}{\partial p_a}+ K^a.
\end{equation}

Then, they proceed to show that, with  $s^\alpha=(\Lambda)^\alpha_{\hphantom{\alpha}\beta}S^\beta$ being the physical spin in the rest frame of an observer along the worldline of the spinning particle and hence $s^\alpha= (0,\mathbf{s})$ being the spin three-vector, equations \eqref{eq:spinequation} and \eqref{eq:velocityequation} are equivalent to the linearized MPDT equations \eqref{eq:linmpd}. Here, $S^\beta$ is the covariant spin vector defined in \eqref{eq:spinvector}, and $(\Lambda)^\alpha_{\hphantom{\alpha}\beta}$ is the Lorentz transformation to the rest frame of the observer along the worldline of the spinning particle. 

\subsection{MPD -- Dirac Equivalence using WKB}\label{sec:diracmpdwkb}

In this section we will give a quick sketch of the derivation of the linearized MPDT equations \eqref{eq:linmpd} from the massive Dirac equation, by using a WKB approximation, along the lines of \cite{rudiger,audretsch}. We start with the Dirac equation \eqref{eq:dirac}:

\begin{equation}
     \gamma^k\nabla_k \Psi - \frac{\mu}{\hbar}\Psi =0,
\end{equation}
with $\mu>0$, and decompose the $4$-spinor $\Psi$ into $2$-spinors:

\[\Psi = \begin{bmatrix}\xi^A\\\overline{\eta}_{A'}\end{bmatrix}. \]

The gamma matrices are given by:

\[\gamma^a = \sqrt{2}\begin{bmatrix}0&-\sigma^{aAB'}\\ \tensor{\sigma}{^a_{BA'}}&0  \end{bmatrix}, \]
and satisfy $\gamma^{(a}\gamma^{b)} = -g^{ab}\mathbbm{1}$. The $\sigma$'s are the Infeld--van der Waerden symbols \cite{PenroseRindler1}, satisfying:

\[\tensor{\sigma}{_a^A_{K'}}\tensor{\sigma}{_b^{BK'}} + \tensor{\sigma}{_b^A_{K'}}\tensor{\sigma}{_a^{BK'}} = g_{ab}\epsilon^{AB}. \]

Once a tetrad is fixed, they become the gamma and Pauli matrices, up to a constant normalization factor. In the following, we will denote the conjugate transpose of a $4$-spinor $\Psi$ by $\Psi^\dagger$, and the Pauli conjugate as: 

\[ \overline{\Psi} =  \begin{bmatrix}\overline{\xi}^{A'} & \eta_A\end{bmatrix}\begin{bmatrix}0&\tensor{\delta}{_{B'}^{A'}}\\ -\tensor{\delta}{_B^A}&0\\ \end{bmatrix} = \begin{bmatrix} -\eta_B& \overline{\xi}^{B'} \end{bmatrix}. \]

Note that $i \overline{\Psi}\Psi$ is real.
Then, one starts the derivation with the WKB ansatz:

\begin{equation}\label{eq:ansatz}
    \Psi = \exp \left(-\frac{i}{\hbar}S\right) \sum_{\nu = 0}^{\infty}\hbar^\nu \psi^{(\nu)},
\end{equation}
where $S$ is a scalar field, and the amplitude $\sum_{\nu = 0}^{\infty}\hbar^\nu \psi^{(\nu)}$ is a $4$-spinor. Plugging the ansatz \eqref{eq:ansatz} into equation \eqref{eq:dirac}, one gets: 

\begin{align}
    (i \gamma^k\nabla_kS + \mu\mathbbm{1})\psi^{(0)} & = 0 &\text{($\nu = 0$)}, \label{eq:ansatzindirac_null}\\
    (i \gamma^k\nabla_kS + \mu\mathbbm{1})\psi^{(\nu)} & = \gamma^k\nabla_k\psi^{(\nu -1)} & \text{($\nu >0$)}.\label{eq:ansatzindirac_nu}
\end{align}

First, we observe that equation \eqref{eq:ansatzindirac_null} has non-zero solution only if $\det(i \gamma^k\nabla_kS + \mu\mathbbm{1}) = 0$. Using $\gamma^{(a}\gamma^{b)} = -g^{ab}\mathbbm{1}$, this leads to:

\begin{equation}\label{eq:solvcond_1}
    \nabla^kS\nabla_kS = \mu^2. \end{equation}

From this equation we can see that geodesics are integral curves of $\frac{1}{\mu}\nabla S$ are. Therefore, up to zero order in $\hbar$, geodesics can be considered a good approximation to solutions of \eqref{eq:dirac}.

We choose an orthonormal frame $e_0, e_1, e_2, e_3,$ with $e_0 = \frac{1}{\mu}\nabla S$, such that the orthonormal frame field is parallel transported along the integral curves of $\frac{1}{\mu}\nabla S$. A spin basis (up to a common sign of $o$ and $\iota$) can now be fixed:

\begin{align}
    o^A\overline{o}^{A'} &= \frac{1}{\sqrt{2}} \tensor{\sigma}{_k^{AA'}}(\tensor{e}{_0^k} + \tensor{e}{_3^k}), \nonumber \\
    \iota^A\overline{\iota}^{A'} &= \frac{1}{\sqrt{2}} \tensor{\sigma}{_k^{AA'}}(\tensor{e}{_0^k} - \tensor{e}{_3^k}), \\
    o^A\overline{\iota}^{A'} &= \frac{1}{\sqrt{2}} \tensor{\sigma}{_k^{AA'}}(\tensor{e}{_1^k} +  i \tensor{e}{_2^k}). \nonumber 
\end{align}

One can then define: 

\begin{equation}
    \Sigma_1 = \frac{1}{\sqrt{2}}\begin{bmatrix}o^A\\ i \overline{\iota}_{A'} \end{bmatrix}, \quad \Sigma_2 = \frac{1}{\sqrt{2}}\begin{bmatrix}\iota^A\\ -i \overline{o}_{A'} \end{bmatrix}, \quad  \Sigma_3 = \frac{1}{\sqrt{2}}\begin{bmatrix}o^A\\ -i \overline{\iota}_{A'} \end{bmatrix}, \quad \Sigma_4 = \frac{1}{\sqrt{2}}\begin{bmatrix}\iota^A\\ i \overline{o}_{A'} \end{bmatrix},
\end{equation}
and calculate: 

\begin{equation}\label{eq:eigenvec}
    (i \gamma^k\nabla_kS + \mu\mathbbm{1})\Sigma_\alpha = \begin{cases}
    0 & \text{for } \alpha = 1,2\\
    2\mu\Sigma_\alpha & \text{for } \alpha = 3,4 \end{cases}.
\end{equation}

Now we impose the following solvability condition:

\begin{equation}\label{eq:solvcond_2}
    \overline{\Sigma}_1\gamma^k\nabla_k\psi^{(\nu -1)} = 0, \quad \overline{\Sigma}_2\gamma^k\nabla_k\psi^{(\nu -1)} = 0.
    \end{equation}
If this condition were not satisfied, \eqref{eq:eigenvec} would prevent any solutions to the equation \eqref{eq:ansatzindirac_nu}.
Then, one can write a general zeroth order solution $\psi^{(0)}$ of \eqref{eq:ansatzindirac_null} as: 

\begin{equation}\label{eq:gensol_null}
    \psi^{(0)} = \tensor{a}{_1^{(0)}}\Sigma_1 + \tensor{a}{_2^{(0)}}\Sigma_2.  \end{equation}

A general solution at order $\nu$ can be given as a sum of the homogeneous solutions as above, and a particular solution $\Pi^{(\nu)}$:

\begin{equation}
    \psi^{(\nu)} = \tensor{a}{_1^{(\nu)}}\Sigma_1 + \tensor{a}{_2^{(\nu)}}\Sigma_2 + \Pi^{(\nu)}.
\end{equation}

The particular solution can be written as $\Pi^{(\nu)} = \tensor{a}{_3^{(\nu)}}\Sigma_3 + \tensor{a}{_4^{(\nu)}}\Sigma_4$. Inserting this into \eqref{eq:ansatzindirac_nu}, using \eqref{eq:eigenvec} on the left hand side, and then equating the coefficients leads to:

\begin{equation}
    \Pi^{(\nu)} = \frac{-i }{2\mu}\left( (\overline{\Sigma}_3\gamma^k\nabla_k\psi^{(\nu -1)})\Sigma_3 + (\overline{\Sigma}_4\gamma^k\nabla_k\psi^{(\nu -1)})\Sigma_4 \right).
\end{equation} 

Finally, the solvability conditions \eqref{eq:solvcond_2} fix all  $\tensor{a}{_1^{(\nu)}}$ and $\tensor{a}{_2^{(\nu)}}$. To obtain the linearized MPDT equations, the authors of \cite{rudiger,audretsch} used the Gordon decomposition, splitting the four-current $j^a = \overline{\Psi}\gamma^a\Psi$ into a convectional current and a spin current:

\begin{equation}\label{eq:gordon_decomp}
    j^a = \underbrace{\frac{\hbar}{2\mu}\left((\nabla^a\overline{\Psi})\Psi - \overline{\Psi}\nabla^a\Psi \right)}_{=: (j_{\text{conv}})^a} + \underbrace{\nabla_k\left( \frac{\hbar}{2\mu}\overline{\Psi}\sigma^{ak}\Psi\right)}_{=:(j_{\text{spin}})^a} ,
\end{equation}
where the relation $\sigma^{ab} = \gamma^{[a}\gamma^{b]}$ has been used.

A comparison to the non-relativistic case shows that the three-current can be interpreted as $j_{\text{conv}}$, and is related to the translation motion of the Dirac particle. The current $j_{\text{spin}}$ is similar to the term one would add to describe the interaction of the magnetic moment of the electron and an external magnetic field. The authors plug the zero-order and first-order terms from \eqref{eq:ansatz} into the definition of $j_{\text{conv}}$ to obtain \cite{rudiger,audretsch}:

\begin{multline}
    (j_{\text{conv}})^a = \frac{1}{\mu}\Big[
    i \overline{\psi}^{(0)}\psi^{(0)}\nabla^aS + \hbar\left(i  \left(\overline{\psi}^{(0)}\psi^{(1)} + \overline{\psi}^{(1)}\psi^{(0)}\right) \right)\nabla^aS \\
    + \frac{1}{2}\left( \left(\nabla^a\overline{\psi}^{(0)}\right)\psi^{(0)} - \overline{\psi}^{(0)}\nabla^a\psi^{(0)} \right) + \dots \Big].
\end{multline}

In the next step, we use the fact that we are concerned with timelike trajectories ($\mu>0$), in order to normalize the current as $u^a := (i \overline{\Psi}\Psi)^{-1}(j_{\text{conv}})^a$. The following definition is introduced:

\[q^a = \frac{1}{2}(i \overline{\Psi}\Psi)^{-1}\left[ \left(\nabla^a\overline{\psi}^{(0)}\right)\psi^{(0)} - \overline{\psi}^{(0)}\nabla^a\psi^{(0)} \right],\] in order to get
\begin{equation}\label{eq:modmom}
    u^a = \mu^{-1}\nabla^aS + \mu^{-1}\hbar q^a .
\end{equation}

Then, one can show that 

\begin{equation}\label{eq:orthog_q}
    q^k\nabla_kS = 0.
\end{equation} 

The following identity is considered

\[ u^k\nabla_ku_a = \frac{1}{2}\nabla_a(u_ku^k) + 2\nabla_{[k}u_{a]}u^k, \]
and on the right hand side we plug in equation \eqref{eq:modmom}. Using \eqref{eq:orthog_q}, the following expression is obtained:

\[\mu(\nabla_ku_a)u^k = \underbrace{\frac{1}{2}\nabla_a\left( (\nabla^kS + \hbar q^k)(\nabla_kS + \hbar q_k) \right)}_{=0} +2\mu\nabla_{[k}u_{a]}u^k .\]
This can be simplified to:

\[ \mu(\nabla_ku_a)u^k = 2\hbar\nabla_{[k}q_{a]}u^k. \]

Now, using $\nabla_{[a}\nabla_{b]}\Psi = -\frac{1}{8}R_{abkl}\sigma^{kl}\Psi$, the authors retrieve equation \eqref{eq:linmpd} as follows \cite{rudiger,audretsch}:

\begin{align}
    \mu(\nabla_ku_a)u^k &= -\frac{1}{4} R_{aklm}
    \underbrace{\hbar(i \overline{\psi}^{(0)}\psi^{(0)})^{-1}\overline{\psi}^{(0)}\sigma^{lm}\psi^{(0)}}_{=2S^{lm}}u^k,\\
    &= -\frac{1}{2}R_{aklm}S^{lm}u^k,
\end{align}
and:

\[ u^k\nabla_kS^{ab} =0.\]

Also, using equation \eqref{eq:ansatzindirac_nu} and its Pauli conjugate, the Pirani-SSC (which to linear order in spin is identical to the Tulczyjew--Dixon SSC) can be obtained \cite{rudiger,audretsch}. Note that $\hbar$ enters here purely as a small expansion parameter, without intrinsic physical meaning.

\section{Discussion}\label{sec:discussion}

We presented various theoretical models attempting to describe G-SHEs. These rely on completely different methods, and the predictions are not entirely consistent. In this section, we will present a final discussion, in order to summarize and compare the main features, limitations and predictions made by each approach. We start with a few comments on the apparent superluminal motion of photons, when subject to the G-SHE, since this can be a common issue, regardless of the method used for deriving the G-SHE.

\subsection{Apparent superluminal motion}

 Several authors \cite{saturnini1976modele, Duval, asenjo2017electromagnetic} reach the conclusion that photons subjected to the G-SHE can move at superluminal speed, despite the fact that photons are generally considered to obey Maxwell's equations, and solutions to Maxwell's equations are known to obey causality. In the following, we want to clarify in what sense these trajectories might actually be valid, or at least to propose an alternative
 interpretation of these results.

First of all, results pertaining to the Gaussian beams in \cite{sbierski2015characterisation} only yield a null geodesic with a Dirac delta initial support, in the limit of $\omega\rightarrow\infty$. However, the G-SHE comes in with a factor of $\sigma/\omega$, and vanishes in the limit mentioned above.

\begin{figure}[t!]
\centering
  \subfloat{%
    \includegraphics[width=.99\textwidth]{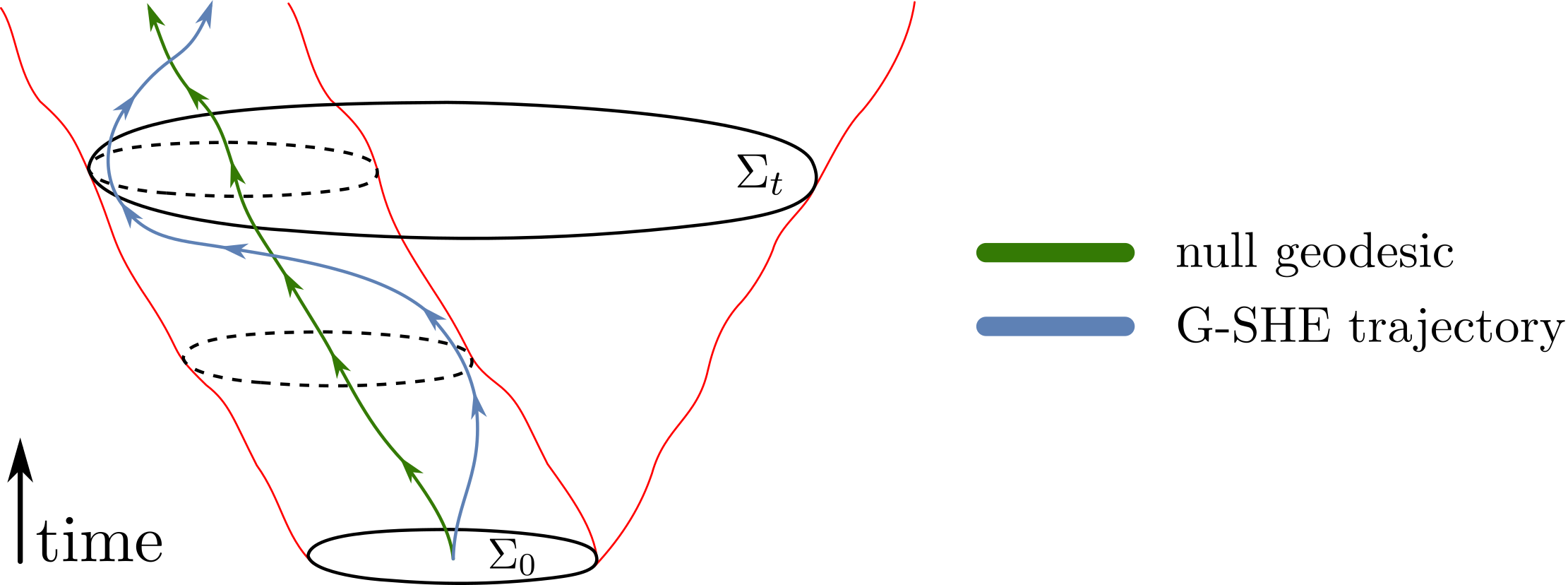}}
\caption{Sketch of a G-SHE trajectory, together with a reference null geodesic. Because the G-SHE disappears for infinite frequencies, we can not assume a solution exhibiting the G-SHE to have a Delta distribution as initial data. Therefore, the information in the initial data is supported in an open region $\Sigma_0$. If the G-SHE trajectory traces the motion of the center of energy, then there is no problem with causality, as long as the G-SHE trajectory remains inside the causal future of the set $\Sigma_0$. This includes the situation where the G-SHE trajectories are spacelike. }
\label{fig:causal}
\end{figure}

This suggests that the solution might lie within an old insight, that, in Minkowski spacetime, one can construct a Gaussian wave packed which has an arbitrary group velocity. Hence, the peak can travel as fast as one wishes. This does not violate the relativistic speed limit, because through the nonvanishing tails of the Gaussian envelop, the information is a priori already present in the entire spacetime. It is likely this phenomenon that is at work here too. If we assume the G-SHE to be present, we cannot take the limit of $\omega\rightarrow\infty$. Hence, we cannot take the initial support of the beam to be arbitrarily small. Parallel transport of the initial data support region, along the original null geodesic to which we consider a modification by the G-SHE gives rise to a tube around the null geodesic. The trajectory calculated for the G-SHE then gives us the path traced out by the center of energy. Within the tube (in fact, within the entire causal future of the support of the initial data), this trajectory can, in principle, be spacelike without any violation of the universal speed limit. A schematic representation can be found in figure \ref{fig:causal}, where $\Sigma_0$ represents the spacelike hypersurface on which the initial data for the electromagnetic wave is supported, and $\Sigma_t$ represents the causal development of $\Sigma_0$ at time $t$.

\subsection{Comparison}

In section \ref{sec:spinningparticles} we presented the MPDT/SouSa approach towards the G-SHE. Looking at the massless case, the SouSa equations have the advantage that, in principle, one can use them to study the G-SHE on any background spacetime. However, the SouSa equations come with some serious drawbacks. First of all, even though they describe massless spinning particles, it is not clear how the SouSa equations are related to Maxwell's equations. Another issue is that certain massless limits of the MPDT equations predict that massless spinning particles will follow null geodesics \cite{mashhoon1975massless, bailyn1977pole, marsot}. In \cite{Duval2018hzh}, the authors analyzed the SouSa equations in Schwarzschild spacetimes, and predicted two different deflection angles for photons passing close to the black hole.

In section \ref{sec:quantum} we discussed the approach of Gosselin, B\'erard and Mohrbach \cite{SHE_QM1,SHE_QM2} towards the G-SHE of light. Here, the authors started with the Bargmann--Wigner equations, and then used the Foldy--Wouthuysen transformation in order to obtain effective equations of motion, describing the G-SHE of photons in a static gravitational field. They predicted a G-SHE for photons travelling in Schwarzschild spacetimes. One important advantage of this method is that the authors started with a field equation, which we consider natural to do when investigating the propagation of light. However, the method is limited to static spacetimes, and it is not clear how one should extend it to more general spacetimes. Also, the physical meaning of the Foldy--Wouthuysen transformation for photons is not clear. When applied to the Dirac equation, the Foldy--Wouthuysen transformation is used to understand the non-relativistic limit. In this sense, it is not clear what a non-relativistic limit of a massless field equation actually means.

In section \ref{sec:maxwell} we presented the modified geometrical optics approach \cite{Frolov,covariantSpinoptics,spinorSpinoptics,spinorSpinoptics2}. Starting with Maxwell's equations on a curved background, the authors used the geometrical optics approximation to derive the gravitational Faraday rotation. Then, using the phase shift associated with the gravitational Faraday rotation, the authors proposed a modified eikonal equation. This lead to the prediction of a G-SHE of light in non-static spacetimes, such as the Kerr spacetime. The main advantage of this method is that one starts with Maxwell's equations, and derives an effect by means of a geometrical optics approach, which is a well known method that has been extensively studied in the literature. However, it is not clear to what extent is the modification of eikonal equation justified. Also, another problem of this approach is that it breaks down at the ergosurface\footnote{For a Kerr black hole, the ergosurface is defined as the region above the outer event horizon where the time translation Killing vector field $\xi = \partial / \partial_t$ becomes null: $g_{\mu \nu} \xi^\mu \xi^\nu = 0$ \cite{visser2007kerr}.} \cite{covariantSpinoptics}.  

Comparing the predictions described in sections \ref{sec:ss} and \ref{sec:quantumpredictions}, we see that the SouSa and the quantum mechanical approach agree to some extend, in the case of a Schwarzschild spacetime. The G-SHE deflection angle predicted by Gosselin et al. \cite{SHE_QM1} is proportional to one of the deflection angles predicted by Duval et al. \cite{Duval2018hzh}. Another important aspect is that the equations of motion discussed in \ref{sec:staticquantum} can also be derived by considering an effective refractive index for the Schwarzschild spacetime, together with the method described in section \ref{sec:SHEL_eq}. This is encouraging, since the theoretical predictions discussed in section \ref{sec:SHEL_eq} have already been tested by several experiments in Optics \cite{SHEL_experiment, Bliokh2008, SOI_review}. 

However, a striking disagreement is seen when one compares the predictions of the quantum mechanical approach with the predictions of the modified geometrical optics approach. As we saw in section \ref{sec:modgo}, the modified geometrical optics approach predicts a G-SHE on Kerr spacetimes, but this effect vanishes in the limit of a Schwarzschild spacetime. This is in contradiction with the predictions presented in sections \ref{sec:ss} and \ref{sec:quantum}, where the G-SHE is present in Schwarzschild spacetimes. Since the methods used for deriving the G-SHE are very different, it is not clear what is the origin of this disagreement. It is necessary to address this issue in future research, so that the theory can make consistent G-SHE predictions, which might one day be measured in experiments.

\section*{Acknowledgements}

While preparing this paper, we discussed the problems at hand with many people to whom we are equally grateful. We will list them here in the alphabetical order. We would like to thank Steffen Acksteiner,  Thomas B\"ackdahl, Iwo Bialynicki-Birula, Ji\v r\'i Bi\v c\'ak, Lukas B\"oke, Ilya Dodin, Sam Dolan, Shane Farnsworth, Abraham Harte, Jan Kohlrus, Miko\l aj Korzy\'nski, Mario Krenn, Andr\'as L\'aszl\'o, Siyuan Ma, Vincent Moncrief, Banibrata Mukhopadhyay, Dennis R\"atzel, Daniel Ruiz, Jan Sbierski, Tomasz Smo\l ka, Justin Vines, and Liviu Z\^arbo for the many useful discussions we had.\\
C.F.P. was supported by the Australian Research Council grant DP170100630. 

\appendix

\section{Specific Spacetimes}
Here, we present some formulas that describe specific spacetimes mentioned in the main text. We omit a discussion of these spacetimes, as it can be found elsewhere (for example, see \cite{griffiths2009}). The relevant properties are mentioned in the main body, where they matter.

The line element for flat Minkowski spacetime, in Cartesian coordinates, is given by:

\begin{equation}\label{eq:minkowski}
    ds^2=-dt^2+dx^2 + dy^2+dz^2.
\end{equation}

In Boyer Lindquist coordinates, $(t, r, \theta, \phi)$, the Schwarzschild metric is given by:

\begin{equation}
ds^2= -\left(1-\frac{2M}{r}\right)dt^2 + \left(1-\frac{2M}{r}\right)^{-1}dr^2 + r^2 d\Omega^2, 
\label{eq:schwarzschild}
\end{equation}
where $d\Omega^2 = d\theta^2 + \sin^2(\theta)d\phi^2$ describes the metric on a unit two-sphere. This describes a one parameter family of solutions, parametrized by $M$. It is straightforward to see that we recover the Minkowski spacetime when we set $M=0$ in this coordinate system. The parameter $M$ is usually interpreted as the mass of the black hole. The Schwarzschild metric is static and asymptotically flat. It can also be expressed in Cartesian isotropic coordinates, $(t, x, y, z)$, in the following way \cite{catalogue_spacetimes}:

\begin{align} \label{eq:isotropic}
    ds^2 = V^2 dt^2 - W^2 (dx^2 + dy^2 + dz^2), \\
    V = \frac{1 - \frac{r_s}{4R}}{1 + \frac{r_s}{4R}}, \qquad W = \left( {1 + \frac{r_s}{4R}} \right)^2, \label{eq:isotropic1}
\end{align}
where $r_s = \frac{2 G M}{c^2}$ is the Schwarzschild radius, and $R = \sqrt{x^2+y^2+z^2}$.   

The Kerr family of spacetimes describes axially symmetric and stationary black hole solutions to the Einstein field equations.  We use Boyer--Lindquist coordinates, ($t, r,\phi,\theta$), which have the property that the metric components are independent of $\phi$ and $t$. The metric has the following form:

\begin{equation}
\label{eq:kerrtaubnut}
\begin{split}
    \mathrm{d}s^2 =& \Sigma \left(\frac{1}{\Delta}\mathrm{d} r^2 + \mathrm{d}\theta^2 \right)+ \frac{1}{\Sigma}\left[(r^2+a^2)^2 \sin^2\theta -\Delta \chi^2\right]\mathrm{d} \phi^2\\
    & -\frac{2}{\Sigma}\left[\Delta\chi-a(r^2+a^2)\sin^2\theta\right]  \mathrm{d}t\mathrm{d}\phi- \frac{1}{\Sigma}\left(\Delta-a^2\sin^2\theta\right)\mathrm{d}t^2,
 \end{split}
\end{equation}
where:
\begin{equation}
\Sigma = r^2 + a^2 \cos^2  \theta,
\end{equation}
\begin{equation}
\chi=a\sin^2(\theta),
\end{equation}
\begin{equation}
\Delta (r)= r^2 - 2Mr + a^2.
\end{equation}
Here, $M$ is the mass, $a$ is the angular momentum per mass unit. 

The spatially-flat Friedmann--Robertson--Walker (FRW) metric  is given by:

\begin{equation}\label{eq:frw}
     \mathrm{d}s^2 = - \mathrm{d}t^2 + {a(t)}^2 \mathrm{d}{\sigma}^2,
\end{equation}
where $\sigma$ is the flat metric in $\Reals^3$, and $a(t)$ is the scale function.

\end{sloppypar}

\newpage

\bibliographystyle{amsplain}
\bibliography{references}

\providecommand{\bysame}{\leavevmode\hbox to3em{\hrulefill}\thinspace}
\providecommand{\MR}{\relax\ifhmode\unskip\space\fi MR }
\providecommand{\MRhref}[2]{%
  \href{http://www.ams.org/mathscinet-getitem?mr=#1}{#2}
}
\providecommand{\href}[2]{#2}
\begin{thebibliography}{100}

\bibitem{Aharonov-Anandan}
Y.~Aharonov and J.~Anandan, \emph{Phase change during a cyclic quantum
  evolution}, Physical Review Letters \textbf{58} (1987), 1593--1596.

\bibitem{Allen92}
L.~Allen, M.~W. Beijersbergen, R.~J.~C. Spreeuw, and J.~P. Woerdman,
  \emph{{Orbital angular momentum of light and the transformation of
  Laguerre-Gaussian laser modes}}, Physical Review A \textbf{45} (1992),
  8185--8189.

\bibitem{classical_Berry2}
J.~Anandan, \emph{Comment on geometric phases for classical field theories},
  Physical Review Letters \textbf{60} (1988), 2555--2555.

\bibitem{Anandan1988}
\bysame, \emph{Non-adiabatic non-abelian geometric phase}, Physics Letters A
  \textbf{133} (1988), no.~4, 171--175.

\bibitem{Berry_CS4}
\bysame, \emph{{Topological and geometrical phases due to gravitational field
  with curvature and torsion}}, Physics Letters A \textbf{195} (1994), no.~5,
  284--292.

\bibitem{AM_Light}
D.~L. Andrews and M.~Babiker, \emph{{The Angular Momentum of Light}}, Cambridge
  University Press, 2012.

\bibitem{asenjo2017electromagnetic}
F.~A. Asenjo and S.~A. Hojman, \emph{Do electromagnetic waves always propagate
  along null geodesics?}, Classical and Quantum Gravity \textbf{34} (2017),
  no.~20, 205011.

\bibitem{audretsch}
J.~Audretsch, \emph{{Trajectories and spin motion of massive spin-$\frac{1}{2}$
  particles in gravitational fields}}, Journal of Physics A: Mathematical and
  General \textbf{14} (1981), 411--422.

\bibitem{bailyn1977pole}
M.~Bailyn and S.~Ragusa, \emph{Pole-dipole model of massless particles},
  Physical Review D \textbf{15} (1977), no.~12, 3543.

\bibitem{bailyn1981pole}
\bysame, \emph{{Pole-dipole model of massless particles. II}}, Physical Review
  D \textbf{23} (1981), no.~6, 1258.

\bibitem{Berry_CS2}
K.~Bakke, C.~Furtado, and J.~R. Nascimento, \emph{{Gravitational geometric
  phase in the presence of torsion}}, The European Physical Journal C
  \textbf{60} (2009), no.~3, 501.

\bibitem{originalSHE3}
A.~A. {Bakun}, B.~P. {Zakharchenya}, A.~A. {Rogachev}, M.~N. {Tkachuk}, and
  V.~G. {Fle{\v i}sher}, \emph{{Observation of a surface photocurrent caused by
  optical orientation of electrons in a semiconductor}}, Soviet Journal of
  Experimental and Theoretical Physics Letters \textbf{40} (1984), 1293.

\bibitem{GW_IOAM4}
P.~Baral, A.~Ray, R.~Koley, and P.~Majumdar, \emph{Gravitational waves with
  orbital angular momentum}, arXiv preprint arXiv:1901.08804 (2019).

\bibitem{barausse2009hamiltonian}
E.~Barausse, E.~Racine, and A.~Buonanno, \emph{{Hamiltonian of a spinning test
  particle in curved spacetime}}, Physical Review D \textbf{80} (2009), no.~10,
  104025.

\bibitem{Analogue_gravity}
C.~Barcel{\'o}, S.~Liberati, and M.~Visser, \emph{{Analogue gravity}}, Living
  Reviews in Relativity \textbf{14} (2011), no.~1, 3.

\bibitem{BargmannWigner_original}
V.~Bargmann and E.~P. Wigner, \emph{{Group theoretical discussion of
  relativistic wave equations}}, Proceedings of the National Academy of
  Sciences \textbf{34} (1948), no.~5, 211--223.

\bibitem{IOAM_electrons4}
S.~M. Barnett, \emph{{Relativistic electron vortices}}, Physical Review Letters
  \textbf{118} (2017), 114802.

\bibitem{AM_Light2}
S.~M. Barnett, L.~Allen, and M.~J. Padgett, \emph{Optical angular momentum},
  CRC Press, 2016.

\bibitem{SHE_QM2}
A.~B\'erard and H.~Mohrbach, \emph{{Spin Hall effect and Berry phase of
  spinning particles}}, Physics Letters A \textbf{352} (2006), no.~3, 190--195.

\bibitem{Berry_original}
M.~V. Berry, \emph{{Quantal phase factors accompanying adiabatic changes}},
  Proceedings of the Royal Society of London A: Mathematical, Physical and
  Engineering Sciences \textbf{392} (1984), no.~1802, 45--57.

\bibitem{Berry-Pancharatnam}
\bysame, \emph{{The adiabatic phase and Pancharatnam's phase for polarized
  light}}, Journal of Modern Optics \textbf{34} (1987), no.~11, 1401--1407.

\bibitem{Pancharatnam-experiment}
R.~Bhandari and J.~Samuel, \emph{{Observation of topological phase by use of a
  laser interferometer}}, Physical Review Letters \textbf{60} (1988),
  1211--1213.

\bibitem{Birula_wavefunction1}
I.~Bialynicki-Birula, \emph{{On the wave function of the photon}}, Acta Physica
  Polonica A \textbf{86} (1994), no.~1, 97--116 (English).

\bibitem{Birula_wavefunction2}
\bysame, \emph{{V Photon wave function}}, Progress in Optics, vol.~36,
  Elsevier, 1996, pp.~245--294.

\bibitem{Maxwell_Berry1}
I.~Bialynicki-Birula and Z.~Bialynicka-Birula, \emph{{Berry's phase in the
  relativistic theory of spinning particles}}, Physical Review D \textbf{35}
  (1987), 2383--2387.

\bibitem{Exponential}
\bysame, \emph{{Exponential beams of electromagnetic radiation}}, Journal of
  Physics B: Atomic, Molecular and Optical Physics \textbf{39} (2006), no.~15,
  S545.

\bibitem{GW_IOAM1}
\bysame, \emph{{Gravitational waves carrying orbital angular momentum}}, New
  Journal of Physics \textbf{18} (2016), no.~2, 023022.

\bibitem{IOAM_electrons5}
\bysame, \emph{{Relativistic electron wave packets carrying angular momentum}},
  Physical Review Letters \textbf{118} (2017), 114801.

\bibitem{GW_IOAM3}
I.~Bialynicki-Birula and S.~Charzy\ifmmode~\acute{n}\else \'{n}\fi{}ski,
  \emph{{Trapping and guiding bodies by gravitational waves endowed with
  angular momentum}}, Physical Review Letters \textbf{121} (2018), 171101.

\bibitem{bini2006massless}
D.~Bini, C.~Cherubini, A.~Geralico, and R.~T. Jantzen, \emph{Massless spinning
  test particles in algebraically special vacuum space--times}, International
  Journal of Modern Physics D \textbf{15} (2006), no.~05, 737--758.

\bibitem{RelativisticQM1}
J.~D. Bjorken and S.~D. Drell, \emph{Relativistic quantum mechanics},
  McGraw-Hill, 1965.

\bibitem{Bliokh2006}
K.~Y. Bliokh, \emph{{Geometrical optics of beams with vortices: Berry phase and
  orbital angular momentum Hall effect}}, Physical Review Letters \textbf{97}
  (2006), 043901.

\bibitem{Bliokh2009}
\bysame, \emph{{Geometrodynamics of polarized light: Berry phase and spin Hall
  effect in a gradient-index medium}}, Journal of Optics A: Pure and Applied
  Optics \textbf{11} (2009), no.~9, 094009.

\bibitem{Bliokh2013}
K.~Y. Bliokh and A.~Aiello, \emph{{Goos–Hänchen and Imbert–Fedorov beam
  shifts: An overview}}, Journal of Optics \textbf{15} (2013), no.~1, 014001.

\bibitem{Bliokh2004}
K.~Y. Bliokh and Y.~P. Bliokh, \emph{{Modified geometrical optics of a smoothly
  inhomogeneous isotropic medium: The anisotropy, Berry phase, and the optical
  Magnus effect}}, Physical Review E \textbf{70} (2004), 026605.

\bibitem{Bliokh2004_1}
\bysame, \emph{{Topological spin transport of photons: The optical Magnus
  effect and Berry phase}}, Physics Letters A \textbf{333} (2004), no.~3,
  181--186.

\bibitem{Bliokh2008}
K.~Y. Bliokh, A.~Niv, V.~Kleiner, and E.~Hasman, \emph{{Geometrodynamics of
  spinning light}}, Nature Photonics \textbf{2} (2008), 748.

\bibitem{lightAM_review}
K.~Y. Bliokh and F.~Nori, \emph{{Transverse and longitudinal angular momenta of
  light}}, Physics Reports \textbf{592} (2015), 1--38.

\bibitem{SOI_review}
K.~Y. Bliokh, F.~J. Rodr\'{i}guez-Fortu\~{n}o, F.~Nori, and A.~V. Zayats,
  \emph{{Spin-orbit interactions of light}}, Nature Photonics \textbf{9}
  (2015), no.~12, 796--808.

\bibitem{Berry_CS5}
A.~Brodutch, T.~F. Demarie, and D.~R. Terno, \emph{{Photon polarization and
  geometric phase in general relativity}}, Physical Review D \textbf{84}
  (2011), 104043.

\bibitem{Berry_CS6}
A.~Brodutch and D.~R. Terno, \emph{{Polarization rotation, reference frames,
  and Mach's principle}}, Physical Review D \textbf{84} (2011), 121501.

\bibitem{Berry_CS1}
Y.~Q. Cai and G.~Papini, \emph{{Applying Berry's phase to problems involving
  weak gravitational and inertial fields}}, Classical and Quantum Gravity
  \textbf{7} (1990), no.~2, 269.

\bibitem{Maxwell_Berry2}
Y.~Q. Cai, G.~Papini, and W.~R. Wood, \emph{{Berry’s phase for photons and
  topology in Maxwell’s theory}}, Journal of Mathematical Physics \textbf{31}
  (1990), no.~8, 1942--1946.

\bibitem{FaradayRotation3}
P.~Carini, L.~L. Feng, M.~Li, and R.~Ruffini, \emph{{Phase evolution of the
  photon in Kerr spacetime}}, Physical Review D \textbf{46} (1992), 5407--5413.

\bibitem{FW_generalization1}
K.~M. Case, \emph{{Some generalizations of the Foldy-Wouthuysen
  transformation}}, Physical Review \textbf{95} (1954), 1323--1328.

\bibitem{chen2015strong}
S.~Chen and J.~Jing, \emph{{Strong gravitational lensing for the photons
  coupled to Weyl tensor in a Schwarzschild black hole spacetime}}, Journal of
  Cosmology and Astroparticle Physics \textbf{2015} (2015), no.~10, 002.

\bibitem{chen2017strong}
S.~Chen, S.~Wang, Y.~Huang, J.~Jing, and S.~Wang, \emph{{Strong gravitational
  lensing for the photons coupled to a Weyl tensor in a Kerr black hole
  spacetime}}, Physical Review D \textbf{95} (2017), no.~10, 104017.

\bibitem{Chiao-Wu}
R.~Y. Chiao and Y.-S. Wu, \emph{{Manifestations of Berry's topological phase
  for the photon}}, Physical Review Letters \textbf{57} (1986), 933--936.

\bibitem{Berry_book}
D.~Chru\'sci\'nski and A.~Jamio\l{}kowski, \emph{{Geometric phases in classical
  and quantum mechanics}}, vol.~36, Springer Science \& Business Media, 2012.

\bibitem{Berry_CS3}
A.~Corichi and M.~Pierri, \emph{Gravity and geometric phases}, Physical Review
  D \textbf{51} (1995), 5870--5875.

\bibitem{costa2012mathisson}
L.~F. Costa, C.~Herdeiro, J.~Nat{\'a}rio, and M.~Zilhao, \emph{{Mathisson’s
  helical motions for a spinning particle: Are they unphysical?}}, Physical
  Review D \textbf{85} (2012), no.~2, 024001.

\bibitem{costa2018spinning}
L.~F.~O. Costa, G.~Lukes-Gerakopoulos, and O.~Semer{\'a}k, \emph{{Spinning
  particles in general relativity: Momentum-velocity relation for the
  Mathisson-Pirani spin condition}}, Physical Review D \textbf{97} (2018),
  no.~8, 084023.

\bibitem{costa2015center}
L.~F.~O. Costa and J.~Nat{\'a}rio, \emph{{Center of mass, spin supplementary
  conditions, and the momentum of spinning particles}}, Equations of Motion in
  Relativistic Gravity, Springer, 2015, pp.~215--258.

\bibitem{costa2016spacetime}
L.~F.~O. Costa, J.~Nat{\'a}rio, and M.~Zilh{\~a}o, \emph{{Spacetime dynamics of
  spinning particles: Exact electromagnetic analogies}}, Physical Review D
  \textbf{93} (2016), no.~10, 104006.

\bibitem{daniels1994faster}
R.~D. Daniels and G.~M. Shore, \emph{{“Faster than light” photons and
  charged black holes}}, Nuclear Physics B \textbf{425} (1994), no.~3,
  634--650.

\bibitem{EM_Schrodinger}
G.~De~Nittis and M.~Lein, \emph{{The Schrödinger formalism of electromagnetism
  and other classical waves — How to make quantum-wave analogies rigorous}},
  Annals of Physics \textbf{396} (2018), 579--617.

\bibitem{dixon1964covariant}
W.~G. Dixon, \emph{{A covariant multipole formalism for extended test bodies in
  general relativity}}, Il Nuovo Cimento (1955-1965) \textbf{34} (1964), no.~2,
  317--339.

\bibitem{dixon1973definition}
\bysame, \emph{{The definition of multipole moments for extended bodies}},
  General Relativity and Gravitation \textbf{4} (1973), no.~3, 199--209.

\bibitem{dixon2015new}
\bysame, \emph{{The new mechanics of Myron Mathisson and its subsequent
  development}}, Equations of Motion in Relativistic Gravity, Springer, 2015,
  pp.~1--66.

\bibitem{Dodin2014}
I.~Y. Dodin, \emph{{Geometric view on noneikonal waves}}, Physics Letters A
  \textbf{378} (2014), no.~22, 1598--1621.

\bibitem{Dodin2018}
I.~Y. Dodin, D.~E. Ruiz, K.~Yanagihara, Y.~Zhou, and S.~Kubo,
  \emph{{Quasioptical modeling of wave beams with and without mode conversion:
  I. Basic theory}}, arXiv preprint arXiv:1901.00268 (2019).

\bibitem{spinorSpinoptics2}
S.~R. Dolan, \emph{{Geometrical optics for scalar, electromagnetic and
  gravitational waves on curved spacetime}}, International Journal of Modern
  Physics D \textbf{27} (2018), 1843010.

\bibitem{spinorSpinoptics}
\bysame, \emph{{Higher-order geometrical optics for circularly-polarized
  electromagnetic waves}}, arXiv preprint arXiv:1801.02273 (2018).

\bibitem{dolgov1998superluminal}
A.~D. Dolgov and I.~D. Novikov, \emph{{Superluminal propagation of light in
  gravitational field and non-causal signals}}, Physics Letters B \textbf{442}
  (1998), no.~1--4, 82--89.

\bibitem{OpticalMagnus}
A.~V. Dooghin, N.~D. Kundikova, V.~S. Liberman, and B.~Ya. Zel'dovich,
  \emph{{Optical Magnus effect}}, Physical Review A \textbf{45} (1992),
  8204--8208.

\bibitem{Duval2013}
C.~Duval, \emph{{Polarized spinoptics and symplectic physics}}, arXiv preprint
  arXiv:1312.4486 (2013).

\bibitem{duval1978conformal}
C.~Duval and H.~H. Fliche, \emph{A conformal invariant model of localized
  spinning test particles}, Journal of Mathematical Physics \textbf{19} (1978),
  no.~4, 749--752.

\bibitem{Duval2006}
C.~Duval, Z.~Horv\'ath, and P.~A. Horv\'athy, \emph{Fermat principle for
  spinning light}, Physical Review D \textbf{74} (2006), 021701.

\bibitem{Duval2007}
\bysame, \emph{{Geometrical spinoptics and the optical Hall effect}}, Journal
  of Geometry and Physics \textbf{57} (2007), no.~3, 925--941.

\bibitem{Duval2018hzh}
C.~Duval, L.~Marsot, and T.~Sch\"ucker, \emph{{Gravitational birefringence of
  light in Schwarzschild spacetime}}, arXiv preprint arXiv:1812.03014 (2018).

\bibitem{Duval}
C.~Duval and T.~Sch\"ucker, \emph{{Gravitational birefringence of light in
  Robertson-Walker cosmologies}}, Physical Review D \textbf{96} (2017), 043517.

\bibitem{SHE_review1}
M.~I. Dyakonov and A.~V. Khaetskii, \emph{{Spin Hall Effect}}, pp.~211--243,
  Springer Berlin Heidelberg, 2008.

\bibitem{originalSHE2}
M.~I. Dyakonov and V.~I. Perel, \emph{{Current-induced spin orientation of
  electrons in semiconductors}}, Physics Letters A \textbf{35} (1971), no.~6,
  459--460.

\bibitem{originalSHE1}
\bysame, \emph{{Possibility of orienting electron spins with current}}, Soviet
  Journal of Experimental and Theoretical Physics Letters \textbf{13} (1971),
  467.

\bibitem{d2016spinning}
G.~d’Ambrosi, S.~Satish~Kumar, J.~van~de Vis, and J.~W. van Holten,
  \emph{Spinning bodies in curved spacetime}, Physical Review D \textbf{93}
  (2016), no.~4, 044051.

\bibitem{d2015covariant}
G.~d’Ambrosi, S.~Satish~Kumar, and J.~W. van Holten, \emph{Covariant
  hamiltonian spin dynamics in curved space--time}, Physics Letters B
  \textbf{743} (2015), 478--483.

\bibitem{Eddington}
A.~S. Eddington, \emph{{Space, time and gravitation: An outline of the general
  relativity theory}}, Cambridge University Press, 1987.

\bibitem{FaradayRotation6}
A.~Farooqui, N.~Kamran, and P.~Panangaden, \emph{{An exact expression for
  photon polarization in Kerr geometry}}, Advances in Theoretical and
  Mathematical Physics \textbf{18} (2014), no.~3, 659--686.

\bibitem{cartographic_analog}
M.~Fathi and R.~T. Thompson, \emph{Cartographic distortions make dielectric
  spacetime analog models imperfect mimickers}, Physical Review D \textbf{93}
  (2016), 124026.

\bibitem{FaradayRotation4}
F.~Fayos and J.~Llosa, \emph{{Gravitational effects on the polarization
  plane}}, General Relativity and Gravitation \textbf{14} (1982), no.~10,
  865--877.

\bibitem{fedorov2013theory}
F.~I. Fedorov, \emph{To the theory of total reflection}, Journal of Optics
  \textbf{15} (2013), no.~1, 014002.

\bibitem{Berry_CS7}
L.-L. Feng and W.~Lee, \emph{Gravitomagnetism and the {Berry} phase of photon
  in an rotating gravitational field}, International Journal of Modern Physics
  D \textbf{10} (2001), no.~06, 961--969.

\bibitem{highOAM}
R.~Fickler, G.~Campbell, B.~Buchler, P.~K. Lam, and A.~Zeilinger, \emph{Quantum
  entanglement of angular momentum states with quantum numbers up to 10,010},
  Proceedings of the National Academy of Sciences \textbf{113} (2016), no.~48,
  13642--13647.

\bibitem{FW_original}
L.~L. Foldy and S.~A. Wouthuysen, \emph{{On the Dirac theory of spin 1/2
  particles and its non-relativistic limit}}, Physical Review \textbf{78}
  (1950), 29--36.

\bibitem{Frolov}
V.~P. Frolov and A.~A. Shoom, \emph{Spinoptics in a stationary spacetime},
  Physical Review D \textbf{84} (2011), 044026.

\bibitem{Frolov2}
\bysame, \emph{Scattering of circularly polarized light by a rotating black
  hole}, Physical Review D \textbf{86} (2012), 024010.

\bibitem{Maxwell_Berry3}
S.~A.~H. Gangaraj, M.~G. Silveirinha, and G.~W. Hanson, \emph{{Berry phase,
  Berry connection, and Chern number for a continuum bianisotropic material
  from a classical electromagnetics perspective}}, IEEE Journal on Multiscale
  and Multiphysics Computational Techniques \textbf{2} (2017), 3--17.

\bibitem{classical_Berry1}
J.~C. Garrison and R.~Y. Chiao, \emph{Geometrical phases from global gauge
  invariance of nonlinear classical field theories}, Physical Review Letters
  \textbf{60} (1988), 165--168.

\bibitem{GH_effect}
F.~Goos and H.~Hänchen, \emph{Ein neuer und fundamentaler versuch zur
  totalreflexion}, Annalen der Physik \textbf{436} (1947), no.~7--8, 333--346.

\bibitem{Gordon}
W.~Gordon, \emph{Zur lichtfortpflanzung nach der relativitätstheorie}, Annalen
  der Physik \textbf{377} (1923), no.~22, 421--456.

\bibitem{Gosselin_diagonalization}
P.~Gosselin, A.~B{\'e}rard, and H.~Mohrbach, \emph{{Semiclassical
  diagonalization of quantum Hamiltonian and equations of motion with Berry
  phase corrections}}, The European Physical Journal B \textbf{58} (2007),
  no.~2, 137--148.

\bibitem{SHE_Dirac}
\bysame, \emph{{Semiclassical dynamics of Dirac particles interacting with a
  static gravitational field}}, Physics Letters A \textbf{368} (2007), no.~5,
  356--361.

\bibitem{SHE_QM1}
\bysame, \emph{{Spin Hall effect of photons in a static gravitational field}},
  Physical Review D \textbf{75} (2007), 084035.

\bibitem{RelativisticQM}
W.~Greiner and D.A. Bromley, \emph{{Relativistic quantum mechanics. Wave
  equations}}, Springer, 2000.

\bibitem{griffiths2009}
J.~B. Griffiths and J.~Podolsk{\`y}, \emph{Exact space-times in {Einstein's}
  general relativity}, Cambridge University Press, 2009.

\bibitem{hackmann2014motion}
E.~Hackmann, C.~L{\"a}mmerzahl, Y.~N. Obukhov, D.~Puetzfeld, and I.~Schaffer,
  \emph{{Motion of spinning test bodies in Kerr spacetime}}, Physical Review D
  \textbf{90} (2014), no.~6, 064035.

\bibitem{SHEL_experiment1}
D.~Haefner, S.~Sukhov, and A.~Dogariu, \emph{{Spin Hall effect of light in
  spherical geometry}}, Physical Review Letters \textbf{102} (2009), 123903.

\bibitem{Haldane1986}
F.~D.~M. Haldane, \emph{Path dependence of the geometric rotation of
  polarization in optical fibers}, Optics Letters \textbf{11} (1986), no.~11,
  730--732.

\bibitem{Haldane1987}
\bysame, \emph{{Comment on ''Observation of Berry's topological phase by use of
  an optical fiber''}}, Physical Review Letters \textbf{59} (1987), 1788--1788.

\bibitem{han2010gravitational}
W.-B. Han, \emph{{Gravitational radiation from a spinning compact object around
  a supermassive Kerr black hole in circular orbit}}, Physical Review D
  \textbf{82} (2010), no.~8, 084013.

\bibitem{Harte2018}
A.~I. Harte, \emph{{Gravitational lensing beyond geometric optics: I. Formalism
  and observables}}, General Relativity and Gravitation \textbf{51} (2019),
  no.~1, 14.

\bibitem{SHEL_experiment}
O.~Hosten and P.~Kwiat, \emph{{Observation of the spin Hall effect of light via
  weak measurements}}, Science \textbf{319} (2008), no.~5864, 787--790.

\bibitem{imbert1972calculation}
C.~Imbert, \emph{Calculation and experimental proof of the transverse shift
  induced by total internal reflection of a circularly polarized light beam},
  Physical Review D \textbf{5} (1972), no.~4, 787.

\bibitem{FaradayRotation5}
H.~Ishihara, M.~Takahashi, and A.~Tomimatsu, \emph{{Gravitational Faraday
  rotation induced by a Kerr black hole}}, Physical Review D \textbf{38}
  (1988), 472--477.

\bibitem{jackson}
J.~D. Jackson, \emph{Classical electrodynamics}, John Wiley \& Sons, 2012.

\bibitem{FW_generalization2}
J.~Jayaraman, \emph{{A note on the recent Foldy-Wouthuysen transformations for
  particles of arbitrary spin}}, Journal of Physics A: Mathematical and General
  \textbf{8} (1975), no.~1, L1.

\bibitem{Kato1950}
T.~Kato, \emph{{On the Adiabatic Theorem of Quantum Mechanics}}, Journal of the
  Physical Society of Japan \textbf{5} (1950), no.~6, 435--439.

\bibitem{originalSHE4}
Y.~K. Kato, R.~C. Myers, A.~C. Gossard, and D.~D. Awschalom, \emph{{Observation
  of the spin Hall effect in semiconductors}}, Science \textbf{306} (2004),
  no.~5703, 1910--1913.

\bibitem{khriplovich1996gravitational}
I.~B. Khriplovich and A.~A. Pomeransky, \emph{Gravitational interaction of
  spinning bodies, center-of-mass coordinate and radiation of compact binary
  systems}, Physics Letters A \textbf{216} (1996), no.~1, 7--14.

\bibitem{kohlrus2018}
J.~Kohlrus, J.~Louko, I.~Fuentes, and D.~E. Bruschi, \emph{{Wigner phase of
  photonic helicity states in the spacetime of the Earth}}, arXiv preprint
  arXiv:1810.10502 (2018).

\bibitem{konstantinov1998superluminal}
M.~Y. Konstantinov, \emph{{Superluminal propagation of light in gravitational
  field and non-causal signals: Some comments}}, arXiv preprint
  arXiv:gr-qc/9810019 (1998).

\bibitem{GW_IOAM2}
M.~Krenn and A.~Zeilinger, \emph{{On small beams with large topological charge:
  II. Photons, electrons and gravitational waves}}, New Journal of Physics
  \textbf{20} (2018), no.~6, 063006.

\bibitem{kyrian2007spinning}
K.~Kyrian and O.~Semer{\'a}k, \emph{{Spinning test particles in a Kerr
  field--II}}, Monthly Notices of the Royal Astronomical Society \textbf{382}
  (2007), no.~4, 1922--1932.

\bibitem{lafrance1995gravity}
R.~Lafrance and R.~C Myers, \emph{{Gravity’s Rainbow: Limits for the
  applicability of the equivalence principle}}, Physical Review D \textbf{51}
  (1995), no.~6, 2584.

\bibitem{IOAM_electrons3}
H.~Larocque, I.~Kaminer, V.~Grillo, G.~Leuchs, M.~J. Padgett, R.~W. Boyd,
  M.~Segev, and E.~Karimi, \emph{{‘Twisted’ electrons}}, Contemporary
  Physics \textbf{59} (2018), no.~2, 126--144.

\bibitem{SHE-L_original}
V.~S. Liberman and B.~Y. Zel'dovich, \emph{{Spin-orbit interaction of a photon
  in an inhomogeneous medium}}, Physical Review A \textbf{46} (1992),
  5199--5207.

\bibitem{Berry_Wigner}
N.~H. Lindner, A.~Peres, and D.~R. Terno, \emph{{Wigner's little group and
  Berry's phase for massless particles}}, Journal of Physics A: Mathematical
  and General \textbf{36} (2003), no.~29, L449.

\bibitem{SHEL_experiment2}
X.~Ling, X.~Yi, X.~Zhou, Y.~Liu, W.~Shu, H.~Luo, and S.~Wen, \emph{{Realization
  of tunable spin-dependent splitting in intrinsic photonic spin Hall effect}},
  Applied Physics Letters \textbf{105} (2014), no.~15, 151101.

\bibitem{SHEL_review}
X.~Ling, X.~Zhou, K.~Huang, Y.~Liu, C.-W. Qiu, H.~Luo, and S.~Wen,
  \emph{{Recent advances in the spin Hall effect of light}}, Reports on
  Progress in Physics \textbf{80} (2017), no.~6, 066401.

\bibitem{SHEL_experiment3}
H.~Luo, X.~Zhou, W.~Shu, S.~Wen, and D.~Fan, \emph{{Enhanced and switchable
  spin Hall effect of light near the Brewster angle on reflection}}, Physical
  Review A \textbf{84} (2011), 043806.

\bibitem{Shadow_deg1}
M.~Mars, C.~F. Paganini, and M.~A. Oancea, \emph{{The fingerprints of black
  holes--shadows and their degeneracies}}, Classical and Quantum Gravity
  \textbf{35} (2018), no.~2, 025005.

\bibitem{marsot}
L.~Marsot, \emph{{How does the photon's spin affect gravitational wave
  measurements?}}, arXiv preprint arXiv:1904.09260 (2019).

\bibitem{mashhoon1975massless}
B.~Mashhoon, \emph{Massless spinning test particles in a gravitational field},
  Annals of Physics \textbf{89} (1975), no.~1, 254--257.

\bibitem{mashhoon2006dynamics}
B.~Mashhoon and D.~Singh, \emph{Dynamics of extended spinning masses in a
  gravitational field}, Physical Review D \textbf{74} (2006), no.~12, 124006.

\bibitem{mathisson2010republication}
M.~Mathisson, \emph{{Republication of: New mechanics of material systems}},
  General Relativity and Gravitation \textbf{42} (2010), no.~4, 1011--1048.

\bibitem{IOAM_electrons2}
B.~J. McMorran, A.~Agrawal, I.~M. Anderson, A.~A. Herzing, H.~J. Lezec, J.~J.
  McClelland, and J.~Unguris, \emph{{Electron vortex beams with high quanta of
  orbital angular momentum}}, Science \textbf{331} (2011), no.~6014, 192--195.

\bibitem{MTW}
C.~W. Misner, K.~S. Thorne, and J.~A. Wheeler, \emph{Gravitation}, W. H.
  Freeman San Francisco, 1973.

\bibitem{Berry_noncyclic2}
N.~Mukunda and R.~Simon, \emph{{Quantum Kinematic Approach to the Geometric
  Phase. I. General Formalism}}, Annals of Physics \textbf{228} (1993), no.~2,
  205--268.

\bibitem{catalogue_spacetimes}
T.~M{\"u}ller and F.~Grave, \emph{Catalogue of spacetimes}, arXiv preprint
  arXiv:0904.4184 (2009).

\bibitem{Murakami2006}
S.~Murakami, \emph{{Intrinsic Spin Hall Effect}}, pp.~197--209, Springer Berlin
  Heidelberg, 2006.

\bibitem{FaradayRotation2}
M.~Nouri-Zonoz, \emph{{Gravitoelectromagnetic approach to the gravitational
  Faraday rotation in stationary spacetimes}}, Physical Review D \textbf{60}
  (1999), 024013.

\bibitem{obukhov2001}
Y.~N. Obukhov, \emph{{Spin, gravity, and inertia}}, Physical Review Letters
  \textbf{86} (2001), 192--195.

\bibitem{obukhov2009}
Y.~N. Obukhov, A.~J. Silenko, and O.~V. Teryaev, \emph{Spin dynamics in
  gravitational fields of rotating bodies and the equivalence principle},
  Physical Review D \textbf{80} (2009), 064044.

\bibitem{obukhov2011}
\bysame, \emph{Dirac fermions in strong gravitational fields}, Physical Review
  D \textbf{84} (2011), 024025.

\bibitem{obukhov2013spin}
\bysame, \emph{Spin in an arbitrary gravitational field}, Physical Review D
  \textbf{88} (2013), no.~8, 084014.

\bibitem{obukhov2017general}
\bysame, \emph{General treatment of quantum and classical spinning particles in
  external fields}, Physical Review D \textbf{96} (2017), no.~10, 105005.

\bibitem{SHE_original}
M.~Onoda, S.~Murakami, and N.~Nagaosa, \emph{Hall effect of light}, Physical
  Review Letters \textbf{93} (2004), 083901.

\bibitem{Palmer2012}
M.~C. Palmer, M.~Takahashi, and H.~F. Westman, \emph{{Localized qubits in
  curved spacetimes}}, Annals of Physics \textbf{327} (2012), no.~4,
  1078--1131.

\bibitem{SHE_pictures}
D.~Pan and H.-X. Xu, \emph{Gravitational field around black hole induces
  photonic spin--orbit interaction that twists light}, Frontiers of Physics
  \textbf{12} (2017), no.~5, 128102.

\bibitem{Pancharatnam1956}
S.~Pancharatnam, \emph{Generalized theory of interference, and its
  applications}, Proceedings of the Indian Academy of Sciences - Section A
  \textbf{44} (1956), no.~5, 247--262.

\bibitem{papapetrou1951spinning}
A.~Papapetrou, \emph{{Spinning test-particles in general relativity. I}},
  Proceedings of the Royal Society of London. Series A, Mathematical and
  Physical Sciences \textbf{209} (1951), no.~1097, 248--258.

\bibitem{Berry_noncyclic3}
A.~K. Pati, \emph{Geometric aspects of noncyclic quantum evolutions}, Physical
  Review A \textbf{52} (1995), 2576--2584.

\bibitem{PenroseRindler1}
R.~Penrose and W.~Rindler, \emph{{Spinors and space-time: Two-spinor calculus
  and relativistic fields}}, vol.~1, Cambridge University Press, 1987.

\bibitem{Perlick2000}
V.~Perlick, \emph{{Ray optics, Fermat’s principle, and applications to
  general relativity}}, vol.~61, Springer Science \& Business Media, 2000.

\bibitem{Perlick2004}
\bysame, \emph{{Gravitational lensing from a spacetime perspective}}, Living
  Reviews in Relativity \textbf{7} (2004), no.~1, 9.

\bibitem{pirani2009republication}
F.~A.~E. Pirani, \emph{{Republication of: On the physical significance of the
  Riemann tensor}}, General Relativity and Gravitation \textbf{41} (2009),
  no.~5, 1215--1232.

\bibitem{Plebansky-Maxwell}
J.~Plebanski, \emph{Electromagnetic waves in gravitational fields}, Physical
  Review \textbf{118} (1960), 1396--1408.

\bibitem{porto2011spin}
R.~A. Porto, A.~Ross, and I.~Z. Rothstein, \emph{{Spin induced multipole
  moments for the gravitational wave flux from binary inspirals to third
  Post-Newtonian order}}, Journal of Cosmology and Astroparticle Physics
  \textbf{2011} (2011), no.~03, 009.

\bibitem{ramirez2015lagrangian}
W.~G. Ram{\'\i}rez and A.~A. Deriglazov, \emph{{Lagrangian formulation for
  Mathisson-Papapetrou-Tulczyjew-Dixon equations}}, Physical Review D
  \textbf{92} (2015), no.~12, 124017.

\bibitem{Ross1984}
J.~N. Ross, \emph{The rotation of the polarization in low birefringence
  monomode optical fibres due to geometric effects}, Optical and Quantum
  Electronics \textbf{16} (1984), no.~5, 455--461.

\bibitem{rudiger}
R.~R{\"u}diger, \emph{{The Dirac equation and spinning particles in general
  relativity}}, Proceedings of the Royal Society of London, Series A,
  Mathematical and Physical Sciences \textbf{377} (1981), 417--424.

\bibitem{Ruiz2017(2)}
D.~E. Ruiz, \emph{A geometric theory of waves and its applications to plasma
  physics}, arXiv preprint arXiv:1708.05423 (2017).

\bibitem{Ruiz2015}
D.~E. Ruiz and I.~Y. Dodin, \emph{First-principles variational formulation of
  polarization effects in geometrical optics}, Physical Review A \textbf{92}
  (2015), 043805.

\bibitem{Ruiz2015(2)}
\bysame, \emph{{Lagrangian geometrical optics of nonadiabatic vector waves and
  spin particles}}, Physics Letters A \textbf{379} (2015), no.~38, 2337--2350.

\bibitem{Ruiz2015(3)}
\bysame, \emph{{On the correspondence between quantum and classical variational
  principles}}, Physics Letters A \textbf{379} (2015), no.~40, 2623--2630.

\bibitem{Ruiz2017}
\bysame, \emph{{Extending geometrical optics: A Lagrangian theory for vector
  waves}}, Physics of Plasmas \textbf{24} (2017), no.~5, 055704.

\bibitem{rytov1938}
S.~M. Rytov, \emph{{On the transition from wave to geometrical optics}},
  Doklady Akademii Nauk SSSR, vol.~18, 1938, pp.~263--267.

\bibitem{Berry_noncyclic1}
J.~Samuel and R.~Bhandari, \emph{{General setting for Berry's phase}}, Physical
  Review Letters \textbf{60} (1988), 2339--2342.

\bibitem{saturnini1976modele}
P.~Saturnini, \emph{{Un modèle de particule à spin de masse nulle dans le
  champ de gravitation}}, Thesis, {Universit{\'e} de Provence}, 1976.

\bibitem{sbierski2015characterisation}
J.~Sbierski, \emph{{Characterisation of the energy of Gaussian beams on
  Lorentzian manifolds: With applications to black hole spacetimes}}, Analysis
  \& PDE \textbf{8} (2015), no.~6, 1379--1420.

\bibitem{Schneiter2018}
F.~Schneiter, D.~Rätzel, and D.~Braun, \emph{{Rotation of polarization in the
  gravitational field of a laser beam - Faraday effect and optical activity}},
  arXiv preprint arXiv:1812.04505 (2018).

\bibitem{semerak1999spinning}
O.~Semer{\'a}k, \emph{{Spinning test particles in a Kerr field—I}}, Monthly
  Notices of the Royal Astronomical Society \textbf{308} (1999), no.~3,
  863--875.

\bibitem{semerak2015spinning}
\bysame, \emph{{Spinning particles in vacuum spacetimes of different curvature
  types: Natural reference tetrads and massless particles}}, Physical Review D
  \textbf{92} (2015), no.~12, 124036.

\bibitem{FaradayRotation1}
M.~Sereno, \emph{{Gravitational Faraday rotation in a weak gravitational
  field}}, Physical Review D \textbf{69} (2004), 087501.

\bibitem{shore1996faster}
G.~M. Shore, \emph{{‘Faster than light’ photons in gravitational
  fields—Causality, anomalies and horizons}}, Nuclear Physics B \textbf{460}
  (1996), no.~2, 379--394.

\bibitem{shore2001accelerating}
\bysame, \emph{Accelerating photons with gravitational radiation}, Nuclear
  Physics B \textbf{605} (2001), no.~1--3, 455--466.

\bibitem{shore2002faster}
\bysame, \emph{{Faster than light photons in gravitational fields II.:
  Dispersion and vacuum polarisation}}, Nuclear Physics B \textbf{633} (2002),
  no.~1--2, 271--294.

\bibitem{shore2003quantum}
\bysame, \emph{Quantum gravitational optics}, Contemporary Physics \textbf{44}
  (2003), no.~6, 503--521.

\bibitem{shore2006causality}
\bysame, \emph{Causality and superluminal light}, Time and Matter, World
  Scientific, 2006, pp.~45--66.

\bibitem{silenko2008foldy}
A.~J. Silenko, \emph{{Foldy-Wouthyusen transformation and semiclassical limit
  for relativistic particles in strong external fields}}, Physical Review A
  \textbf{77} (2008), no.~1, 012116.

\bibitem{Silenko2005}
A.~J. Silenko and O.~V. Teryaev, \emph{{Semiclassical limit for Dirac particles
  interacting with a gravitational field}}, Physical Review D \textbf{71}
  (2005), 064016.

\bibitem{Berry_Simon}
B.~Simon, \emph{{Holonomy, the quantum adiabatic theorem, and Berry's phase}},
  Physical Review Letters \textbf{51} (1983), 2167--2170.

\bibitem{singh2008analytic}
D.~Singh, \emph{An analytic perturbation approach for classical spinning
  particle dynamics}, General Relativity and Gravitation \textbf{40} (2008),
  no.~6, 1179--1192.

\bibitem{singh2008perturbation}
\bysame, \emph{{Perturbation method for classical spinning particle motion. I.
  Kerr space-time}}, Physical Review D \textbf{78} (2008), no.~10, 104028.

\bibitem{Sinitsyn2007}
N.~A. Sinitsyn, \emph{{Semiclassical theories of the anomalous Hall effect}},
  Journal of Physics: Condensed Matter \textbf{20} (2007), no.~2, 023201.

\bibitem{SHE_review}
J.~Sinova, S.~O. Valenzuela, J.~Wunderlich, C.~H. Back, and T.~Jungwirth,
  \emph{{Spin Hall effects}}, Reviews of Modern Physics \textbf{87} (2015),
  1213--1260.

\bibitem{NC_Maxwell}
S.~Sivasubramanian, G.~Castellani, N.~Fabiano, A.~Widom, J.~Swain, Y.~N.
  Srivastava, and G.~Vitiello, \emph{Non-commutative geometry and measurements
  of polarized two photon coincidence counts}, Annals of Physics \textbf{311}
  (2004), no.~1, 191--203.

\bibitem{SHEL_experiment4}
A.~P. Slobozhanyuk, A.~N. Poddubny, I.~S. Sinev, A.~K. Samusev, Y.~F. Yu, A.~I.
  Kuznetsov, A.~E. Miroshnichenko, and Y.~S. Kivshar, \emph{{Enhanced photonic
  spin Hall effect with subwavelength topological edge states}}, Laser \&
  Photonics Reviews \textbf{10} (2016), no.~4, 656--664.

\bibitem{souriau1974modele}
J.-M. Souriau, \emph{Modele de particulea spin dans le champ
  {\'e}lectromagn{\'e}tique et gravitationnel}, Annales de l'Institut Henri
  Poincaré A \textbf{20} (1974), 315--364.

\bibitem{steinhoff2015spin}
J.~Steinhoff, \emph{Spin and quadrupole contributions to the motion of
  astrophysical binaries}, Equations of Motion in Relativistic Gravity,
  Springer, 2015, pp.~615--649.

\bibitem{Tomita-Chiao}
A.~Tomita and R.~Y. Chiao, \emph{{Observation of Berry's topological phase by
  use of an optical fiber}}, Physical Review Letters \textbf{57} (1986),
  937--940.

\bibitem{tulczyjew1959motion}
W.~Tulczyjew, \emph{Motion of multipole particles in general relativity
  theory}, Acta Physica Polonica \textbf{18} (1959), 393.

\bibitem{IOAM_electrons1}
M.~Uchida and A.~Tonomura, \emph{Generation of electron beams carrying orbital
  angular momentum}, Nature \textbf{464} (2010), 737.

\bibitem{van2016spinning}
J.~W. van Holten, \emph{Spinning bodies in general relativity}, International
  Journal of Geometric Methods in Modern Physics \textbf{13} (2016), no.~08,
  1640002.

\bibitem{van2016world}
\bysame, \emph{World-line perturbation theory}, pp.~393--418, Springer
  International Publishing, Cham, 2019.

\bibitem{vines2016canonical}
J.~Vines, D.~Kunst, J.~Steinhoff, and T.~Hinderer, \emph{{Canonical Hamiltonian
  for an extended test body in curved spacetime: To quadratic order in spin}},
  Physical Review D \textbf{93} (2016), no.~10, 103008.

\bibitem{visser2007kerr}
M.~Visser, \emph{{The Kerr spacetime: A brief introduction}}, arXiv preprint
  arXiv:0706.0622 (2007).

\bibitem{vladimirskii1941}
V.~V. Vladimirskii, \emph{{The rotation of polarization plane for curved light
  ray}}, Doklady Akademii Nauk SSSR, vol.~31, 1941, pp.~222--226.

\bibitem{Bessel}
K.~Volke-Sepulveda, V.~Garcés-Chávez, S.~Chávez-Cerda, J.~Arlt, and
  K.~Dholakia, \emph{{Orbital angular momentum of a high-order Bessel light
  beam}}, Journal of Optics B: Quantum and Semiclassical Optics \textbf{4}
  (2002), no.~2, S82.

\bibitem{wang2016effect}
Z.-Y. Wang, C.-D. Xiong, Q.~Qiu, Y.-X. Wang, and S.-J. Shi, \emph{{The effect
  of gravitational spin--orbit coupling on the circular photon orbit in the
  Schwarzschild geometry}}, Classical and Quantum Gravity \textbf{33} (2016),
  no.~11, 115020.

\bibitem{Wilczek-Zee}
F.~Wilczek and A.~Zee, \emph{Appearance of gauge structure in simple dynamical
  systems}, Physical Review Letters \textbf{52} (1984), 2111--2114.

\bibitem{Berry_electronic}
D.~Xiao, M.-C. Chang, and Q.~Niu, \emph{Berry phase effects on electronic
  properties}, Reviews of Modern Physics \textbf{82} (2010), 1959--2007.

\bibitem{SHE_GW}
N.~Yamamoto, \emph{{Spin Hall effect of gravitational waves}}, Physical Review
  D \textbf{98} (2018), 061701.

\bibitem{covariantSpinoptics}
C.-M. Yoo, \emph{Notes on spinoptics in a stationary spacetime}, Physical
  Review D \textbf{86} (2012), 084005.

\end{thebibliography}

\end{document}